\documentclass[a4paper, 11pt]{article}

\usepackage{graphicx}
\usepackage{amsmath} % Advanced mathematical typesetting
\usepackage{calc} % Used to position the abstract
\usepackage[hmargin=3cm, vmargin=2.5cm]{geometry} % Tweaks the margins
\usepackage[hidelinks]{hyperref} % All references work as clickable links

\usepackage{mathpazo} 
\usepackage{helvet} % Sets the sans-serif font to a Helvetica copy
\usepackage{bm} % Bold italic math for vectors

\usepackage[font={sf, small}]{caption}
\usepackage[sf, bf, medium]{titlesec}

\usepackage[square,comma,sort&compress,numbers]{natbib}
\usepackage{fancyhdr}
\begin{document}

% Title and author list
\null
\begin{center}
\sffamily \LARGE \bfseries

On parabolic-equation model verification for underwater acoustics \\ % ENTER THE PAPER TITLE HERE

\normalsize \mdseries \vspace{\baselineskip}

Sven M. Ivansson\textsuperscript{1} \\ % AUTHOR 1
% Feel free to add more authors as necessary, of course

\footnotesize \vspace{\baselineskip}

\textsuperscript{1} Stockholm, SE-11529, Sweden \\ % AFFILIATION 1
% Feel free to add more affiliations as necessary, of course
\newcommand{\email}{sven.ivansson@gmail.com} % EMAIL OF CONTACT AUTHOR
Contact email: \href{mailto:\email}{\email}

\end{center}
\vspace{\baselineskip}

% Abstract
\noindent\hspace*{\fill}\begin{minipage}{\textwidth-2cm}
{\Large \sffamily \centering \textbf{Abstract} \\} \small

% FILL OUT THIS ABSTRACT
Modeling of sound propagation in media with azimuthal
variations of the material parameters typically
necessitates approximations. Useful methods, such as
the 3-D parabolic equation (PE) method, need verification
by correct reference solutions for well-defined test
examples.                                               %  34
The present paper utilizes reference
solutions for media with particular types of lateral
sound-speed variation, as a complement to common
wedge and canyon examples with bathymetry variation.     % 31
Notably, the adiabatic approximation is exact for the particular media
under study, implying, however, that verification of
mode coupling remains.                                   % 33
Wavenumber integration, for computation of modal
expansion coefficients, produces accurate
solutions for the field and its spatial derivatives in the 3-D case. 
Explicit expressions are typically available
for the wavenumber integrands in terms of Airy and exponential
functions. For a related 2-D case with azimuthal symmetry,
Hankel functions appear instead of exponentials, and
the wavenumber integration drops out.                     % 38
The PE verification focuses on comparing 
the two sides of the PE at insertion of appropriately
scaled Helmholtz-equation solutions.
\end{minipage}\hspace*{\fill}
\vspace{\baselineskip}

%------------------------------------------------------------------------------
\section{Introduction}

Parabolic equation (PE) modeling, with replacement of the
second-order Helmholtz equation by an equation
that is first-order, one-way, in range, 
greatly simplifies sound propagation modeling for
laterally varying environments~\cite{jensen11}. 
However, the approximation necessitates 
verification and bench-marking, involving
accurate reference solutions by other methods
for well-defined test cases.

Starting with a paper by Evans~\cite{evans83}, 
full coupled-mode computations have been used
extensively for 2-D model verification.
Concerning 3-D modeling, 
wedge and canyon examples, with restriction of the
medium variation to one of the two Cartesian
horizontal coordinates, have been popular for
the model verification~\cite{fawcett93,sturm05}. 
The present paper utilizes accurate 
reference solutions for related types of
media, with lateral variation of sound speed
rather than bathymetry.
In principle, a PE has the form
\begin{equation} \label{eq:pe}
  \frac{\partial \psi}{\partial r} = U(\psi)
\end{equation}
where $\psi$ is acoustic pressure $p$, scaled in a
certain way, $r$ is range, and $U$ is some spatial
differential operator not including derivatives
with respect to $r$.
The typical verification approach
is to assess the difference between the $\psi$ 
that solves Eq.~(\ref{eq:pe}) and the $\psi$ that 
corresponds to the underlying Helmholtz-equation
solution for $p$. 
The present paper, however, focuses the difference between the
two sides of Eq.~(\ref{eq:pe}) when the latter $\psi$ is 
inserted. 

When the material medium parameters depend on only one
of the two Cartesian horizontal coordinates, 
Fourier transformation with respect to the other
horizontal coordinate recasts the 3-D Helmholtz-equation problem
as an ensemble of 2-D problems~\cite{fawcett90}. 
If, in addition, neither the difference of
inverse squared sound speeds nor the quotient of densities
at two arbitrary horizontal positions depend on depth, 
it follows that the local modes are the same at all 
horizontal positions~\cite[Sec. 7.1.2]{brekh92}. 
Only the modal wavenumbers differ, and the adiabatic
approximation is exact. 

Section~2 concerns a related 2-D case, 
for an azimuthally symmetric medium.
Reflection back towards the source, rather than
penetration into the bottom, turns out to be a possible
mechanism for mode cutoff.
Section~3 concerns the 3-D case, with an accurate
wavenumber integration method to determine 
the lateral variation of pertinent adiabatic-mode
expansion coeefficients.
Continuous-wave as well as broad-band examples are
included. Accompanying
computations of energy flux and horizontal rays
are helpful for an intuitive understanding of the
field results. 
PE-method verification examples appear in Secs.~2.2 and~3.3. 
The final Sec.~4 contains some concluding remarks.
Parts of Sec.~3 have previously been
presented in~\cite{ivan17b}.

%------------------------------------------------------------------------------
\section{The 2-D case}
% se ``11/9 17'' blad 1-2,4-8

In a Cartesian $xyz$ coordinate system, with horizontal coordinates
$x$,$y$ and depth coordinate $z$, the Helmholtz equation for the acoustic
pressure $p$ in a fluid medium with density $\rho = \rho(x,y,z)$ and
sound speed $c = c(x,y,z)$ appears as~\cite[Eq.~(3.9)]{ivan17a} 
\begin{equation} \label{eq:helm}
  \bigtriangleup p + \rho \cdot \nabla (\rho^{-1}) \cdot \nabla p +
  \frac{\omega^{2}}{c^{2}} p = \frac{\omega^{2}}{c^{2}} M \delta_{s} \ . 
\end{equation}
Here, $\omega > 0$ and $M$ denote the angular frequency and the
momont-tensor strength, respectively, of a symmetric point source at
$\bm{x}_{s} = (x_{s},y_{s},z_{s})$. 
Furthermore, $\bigtriangleup = \partial^{2}/\partial x^{2} +
\partial^{2}/\partial y^{2} + \partial^{2}/\partial z^{2}$ is the Laplacian, 
$\nabla = (\partial/\partial x,\partial/\partial y,\partial/\partial z)$
is the gradient operator, and $\delta_{s}$ is the three-dimensional Dirac delta 
function $\delta$ centered at the source (\textit{i.e.}, 
$\delta_{s}(\bm{x}) = \delta_{s}(x,y,z) = \delta(\bm{x}-\bm{x}_{s})$). 

Next, introduce polar coordinates $(r,\theta)$ in the horizontal $xy$-plane. 
When the medium parameters do not depend on the azimuthal angle $\theta$,
\textit{i.e.}, when $\rho = \rho(r,z)$ and $c = c(r,z)$, and
when $(x_{s},y_{s})$ = (0,0) m, the Helmholtz equation~(\ref{eq:helm})
appears as 
\begin{equation} \label{eq:helm_2d}
  \frac{1}{r} \frac {\partial} {\partial r}
    \left( r \frac {\partial p} {\partial r} \right)
    + \rho \frac {\partial} {\partial r}  \left( \frac {1} {\rho} \right)
       \frac {\partial p} {\partial r} 
    + \rho \frac {\partial} {\partial z}  \left( \frac {1} {\rho} \, 
       \frac {\partial p} {\partial z} \right) 
    + \frac{\omega^{2}}{c^{2}} p 
  = \frac{\omega^{2}}{c^{2}} M \frac {\delta(r) \delta(z-z_{s}) } {2 \pi r}
\end{equation}
in $rz$ coordinates, where $\delta$ is now the one-dimensional Dirac delta 
function. 

Consider now particular media such that 
\begin{eqnarray}
  \rho(r,z) & = & \rho_{0}(z) \, R(r) \label{eq:rho0_2d} \\
  \frac{1}{c^{2}(r,z)} & = & \frac{1}{c_{0}^{2}(z)} + S(r) \ , 
                                                         \label{eq:c0_2d}
\end{eqnarray}
where $R(0)=1$ and $S(0)=0$ s$^{2}$/m$^{2}$. 
Hence, $\rho_{0}(z)$ and $c_{0}(z)$ denote the density and sound-speed
profiles, respectively, at $r=0$.
It is convenient to restrict $R(r)>0$ and $S(r)$ to be 
constant functions for large as well as small positive $r$. 
There is a free boundary at depth $z=0$ and a free or rigid
boundary at depth $z_{h}$. 
At $r$ = 0 m, $Z_{m,0}(z)$, $m$ = 1,2,.., form a complete set of orthogonal
(with respect to the weight function $\rho_{0}^{-1}(z)$) 
local modes with corresponding modal wavenumbers $k_{m,0}$. 
With an underlying dependence on time $t$ according to
$\exp(-i \omega t)$, the modal wavenumbers are 
in the upper complex half-plane and, when they are real, on the positive 
rather than negative real axis.
Winding-number techniques are useful to determine the
$k_{m,0}$ (\textit{e.g.},~\cite[Sec. 3.3.1]{ivan17a}), 
and it is appropriate to order the modal wavenumbers according to 
decreasing real parts. 
The ordinary differential equations for $Z_{m,0}(z)$ can be solved
conveniently by 2$\times$2 propagator-matrix techniques as
detailed in~\cite[Sec. 3.3.2]{ivan17a}, for example.
At range $r$, it follows that the $Z_{m,0}(z)$ are local modes with
modal wavenumbers
$ k_{m}(r) = ( k_{m,0}^{2} + \omega^{2} S(r) )^{1/2} $. 

Expansion of the pressure $p$ in Eq.~(\ref{eq:helm_2d}) in terms 
of the $Z_{m,0}$, \textit{i.e.}, 
\begin{equation} \label{eq:ansatz_2d}
  p(r,z) = \sum_{m=1}^{\infty} \gamma_{m}(r) \, Z_{m,0}(z)  \ , 
\end{equation}
provides the ordinary Helmholtz-type differential equations 
\begin{equation} \label{eq:hrefract_2d}
  \frac{R(r)}{r} \frac{d}{d r}
    \left( \frac{r}{R(r)} \frac{d \gamma_{m}(r)}{d r} \right)  +
    k_{m}^{2}(r) \gamma_{m}(r) 
  = \frac{\omega^{2}}{c^{2}(0,z_{s})} M \frac{\delta(r)}{2 \pi r} 
    \frac {Z_{m,0}(z_{s})} { I_{m,0} \, \rho_{0}(z_{s}) } 
\end{equation}
for the coefficient functions $\gamma_{m}(r)$, where $m$ = 1,2,.. and 
\begin{equation} \label{eq:Idef}
   I_{m,0} = \int_{0}^{z_{h}} \frac{1}{\rho_{0}} Z_{m,0}^{2} \, dz \ . 
\end{equation}
There is no mode coupling and the
adiabatic approximation is exact. 
With $R^{-1/2}(r) \, \gamma_{m}(r)$ as dependent variable, 
Eq.~(\ref{eq:hrefract_2d}) agrees with~\cite[Eq. (3.102)]{ivan17a}. 

In a subinterval of (0,+$\infty$) where $R(r)$ and $S(r)$ are
constant, $\gamma_{m}(r)$ is obviously a linear combination of
the Hankel functions
$H_{0}^{(1)}(k_{m}(r)r)$ and $H_{0}^{(2)}(k_{m}(r)r)$.
In the outermost subinterval, out to $+\infty$, 
$ \gamma_{m}(r) = \gamma_{m}^{+} \, H_{0}^{(1)}(k_{m}(r)r) $, 
where $\gamma_{m}^{+}$ is a coefficient to be determined by 
propagation inwards of 
$ ( (R \omega^{2})^{-1} \, d \gamma_{m} / d r , -\gamma_{m} )^{\mathrm{T}} $,
\textit{cf.}~\cite[Sec. 3.2.1.2]{ivan17a}. 
In the innermost subinterval, with $r$=0 m to the left,
the source condition of Eq.~(\ref{eq:hrefract_2d}) implies 
that 
\begin{equation} \label{eq:solv0}
  \gamma_{m}(r)  =  - i \frac {\omega^{2}} {4 c^{2}(0,z_{s}) } M 
    \frac {Z_{m,0}(z_{s})} { I_{m,0} \, \rho_{0}(z_{s}) } \, 
    H_{0}^{(1)}(k_{m}(r)r) + \, \gamma_{m}^{-} \, J_{0}(k_{m}(r)r)  \ . 
\end{equation}
Solution matching of $\gamma_{m}(r)$ and $d \gamma_{m}(r) / d r$
at the right end of this innermost interval determines 
the coefficients $\gamma_{m}^{+}$ and $\gamma_{m}^{-}$. 
Obviously, the computation method also yields the 
first-order derivatives $ d \gamma_{m}(r) / d r $, 
and higher-order derivatives follow readily by 
differentiation of Eq.~(\ref{eq:hrefract_2d}). 

Analytic solutions of Eq.~(\ref{eq:hrefract_2d}) also appear
in subintervals $(r_{1},r_{2})$ of (0,+$\infty$) where 
$R(r)$ is constant and
$ S(r) + 1/4\omega^{2}r^{2} $ is a linear function of $r$. 
For such a subinterval, $ r^{1/2} \, \gamma_{m}(r) $ is a 
simple linear combination of Airy functions. 
In the exceptional case when
$ S(r) + 1/4\omega^{2}r^{2} $ is constant, 
the Airy functions degenerate to ordinary exponentials.

%-------------------------------------------------------------------
\subsection{\textit{Example mimicking a 2-D wedge}}

Consider a Pekeris wave-guide with a 200 m deep homogeneous water column
overlying a homogeneous fluid sediment half-space with
density 1500 kg/m$^{3}$. 
At the position $(x_{s},y_{s})$ = (0,0) m for a symmetric point source,
the water and sediment sound speeds are 1500 and 3000 m/s, respectively,
and the sediment absorption is 0.1 dB/wavelength. 
For technical convenience, the sediment is truncated by a free boundary
at depth $z$ = 1450 m, and there is a gradual absorption increase below
$z$ = 450 m, from 0.1 to 10 dB/wavelength, to effectively reduce reflections
from this boundary. 
At 30 Hz, there are seven propagating local modes, with real part of
the modal slowness exceeding 1/3000 s/m. (A large sediment sound speed is
useful for illustration clarity, to allow modes with clearly different
slownesses.) For the source depth $z_{s}$ = 106 m, modes 2, 4, and 6 are
very weak~\cite{ivan17b}. 

With azimuthal symmetry according to Eqs.~(\ref{eq:rho0_2d}) 
and~(\ref{eq:c0_2d}), $S(r)$ now prescribes an increase of
the ``water'' sound speed from 1500 m/s for $r \leq 3$ km to
1900 m/s for $r \geq 4$ km. 
The implied ``sediment'' sound speed and absorption beyond
$r$ = 4 km are very large. 
It is convenient here to keep the water-column and sediment-layer terminology, 
although the sound-speed values indicate a gradual lateral change from
water to sediment. 
To allow analytic solutions in terms of Airy (together with Hankel)
functions, $ S(r) + 1/4\omega^{2}r^{2} $ is linear 
between $r$ = 3 and 4 km. 
For simplicity, $R(r)=1$ identically. 

\begin{figure}[t]
\begin{center}
\vspace*{-2.0cm}
\includegraphics [width=7.0cm] {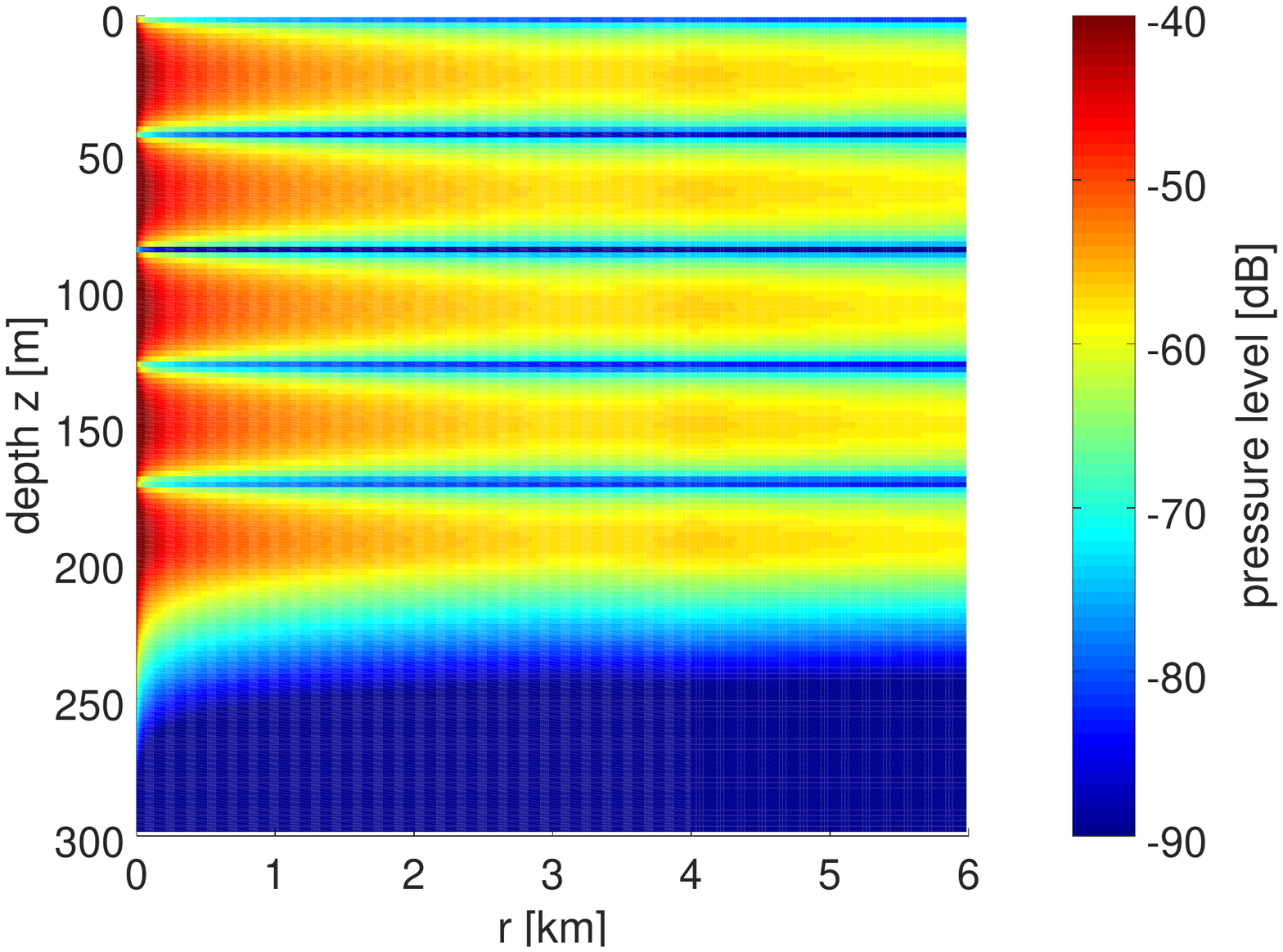}
\includegraphics [width=7.0cm] {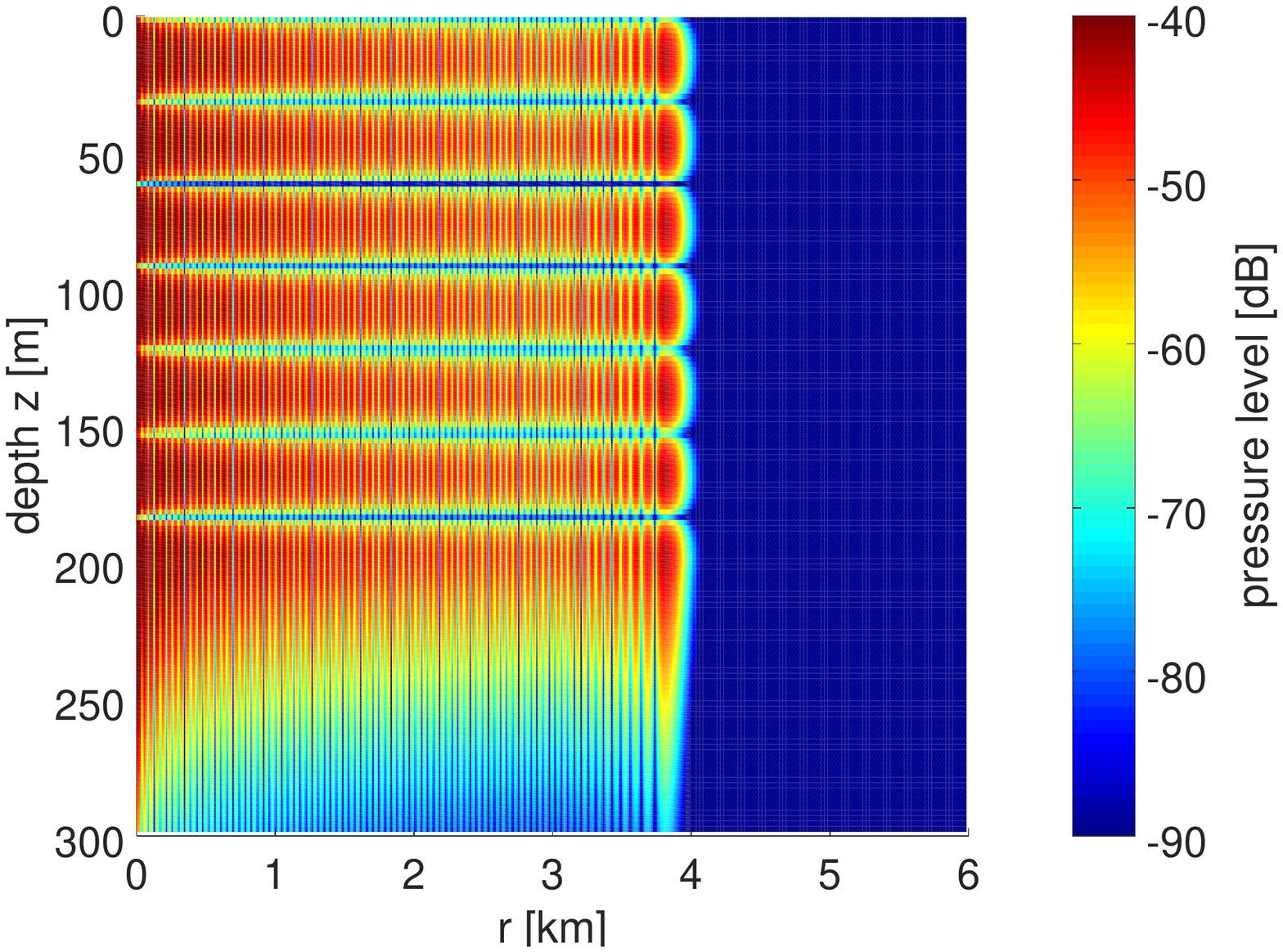}
\vspace*{-2cm}
\caption{ Pressure level, in dB \textit{re}
  total spherical field at 1 m, of the field component given
  by mode 5 (left panel) and 7 (right panel), for the 2-D example, in the $rz$-plane.
  The symmetric source is at depth $z_{s}$ = 106 m
  and the frequency is 30 Hz. }
\label{fig:RZ6_5_7}
\end{center}
\end{figure}

Figure~\ref{fig:RZ6_5_7} shows pressure levels, 
in dB \textit{re} total spherical field at 1 m, in the horizontal
$rz$-plane for the mode 5 and mode 7 field components. 
Mode 5 propagates through the 
sound-speed increase region between $r$ = 3 and 4 km,
with a magnitude peak at about $r$ = 4 km, 
while mode 7 is cut off close to this range. 
In general terms, cutoff for mode $m$ appears when 
$ k_{m}^{2}(r) = k_{m,0}^{2} + \omega^{2} S(r) $
enters the left half-plane and, for a subsequent range-invariant 
interval, the modal Hankel-function factor
$H_{0}^{(1)}(k_{m}(r)r)$ governs propagation with
significant exponential decay. 
This happens for mode 7 when the ``water'' sound-speed
reaches 1837 m/s, \textit{i.e.}, just before $r$ = 4 km,
while it never happens for mode 5 with its larger (real
part of the) modal wavenumber $k_{m,0}$. 

%--------------------------------------------------------
%start Fig. RZ6_5_7; 
%--------------------------------------------------------

Moreover, the mode 7 field exhibits a standing-wave
interference pattern that is absent for mode 5. At cutoff,
the modal energy turns back towards the source,
with a focusing effect in this azimuthally symmetric
environment. Hence, constructive as well as destructive
interference arises between the out-going and in-coming
field components. In the range-invariant $r < $ 3 km region,
these field components are governed (with $m$=7) by the
modal Hankel-function factors $H_{0}^{(1)}(k_{m}(r)r)$ and
$H_{0}^{(2)}(k_{m}(r)r)$, respectively.
Closer to $r$ = 4 km, where $k_{m}(r)$ is smaller,
the period of the interference pattern is larger. 

It is instructive to compare to the classical wedge examples
by Jensen and Kuperman~\cite{jensen80}, with mode cutoff
at upslope propagation in a homogeneous water column. 
Mode penetration into the bottom, rather than reflection back 
towards the source, showed up at cutoff in that study, 
which involved PE modeling and comparisons with tank
experiments.

%-------------------------------------------------------------------
\subsection{\textit{2-D PE approximation errors}}
% se ``27/9 17''

For the cylindrically symmetric case, when the medium parameters
do not depend on the polar $\theta$ coordinate, 
the 2-D Godin PE~\cite{godin99} for
$ \psi(r,z) = p(r,z) \, / \, H_{0}^{(1)}(k_{0}r)$, 
\textit{cf.}~\cite[Eq.~(3.130)]{ivan17a}, appears as 
\begin{equation} \label{eq:pe2d}
  \frac
     { \partial \left( \rho^{-1/2} (I+X)^{1/4} \, \psi \right) }
     { \partial r } = i k_{0} \, \rho^{-1/2} \left( (I+X)^{1/2} - I \right) 
            (I+X)^{1/4} \, \psi \, . 
\end{equation}
It is valid outside the source
at $r$ = 0 m. $I$ is the identity operator, 
$ X = k_{0}^{-2} \, ( \, \rho \, 
  \partial (\rho^{-1} \partial / \partial z) / \partial z +
  ( \omega^{2}/c^{2} - k_{0}^{2} ) \, ) $, and
$k_{0}$ is a reference wavenumber. 
With medium parameters according to Eqs.~(\ref{eq:rho0_2d})-~(\ref{eq:c0_2d}), 
and the local modes $Z_{m,0}(z)$ with modal wavenumbers 
$ k_{m}(r) = ( k_{m,0}^{2} + \omega^{2} S(r) )^{1/2} $, 
$ (I+X) (Z_{m,0}) = (k_{m}/k_{0})^{2} \, Z_{m,0} $. 
Hence, it is natural to define the operators $(I+X)^{1/2}$ and
$(I+X)^{1/4}$ by
$ (I+X)^{1/2} (Z_{m,0}) = (k_{m}/k_{0}) Z_{m,0} $ and 
$ (I+X)^{1/4} (Z_{m,0}) = (k_{m}/k_{0})^{1/2} \, Z_{m,0} $, 
respectively. 
With the mode expansion 
$ \psi(r,z) = \sum_{m=1}^{\infty} \eta_{m}(r) \, Z_{m,0}(z) $,
\textit{cf.} Eq.~(\ref{eq:ansatz_2d}), 
the PE~(\ref{eq:pe2d}) takes the form 
\begin{equation} \label{eq:pe2d_eta}
  \frac { d ( R^{-1/2} (k_{m}/k_{0})^{1/2} \, \eta_{m} ) } { d r } = i k_{0} 
  ( k_{m}/k_{0} - 1 ) R^{-1/2} (k_{m}/k_{0})^{1/2} \, \eta_{m} \ , 
\end{equation}
where $m$ = 1,2,... 
Again, there is no mode coupling. 

\begin{figure}[t]
\begin{center}
\includegraphics [width=7.0cm] {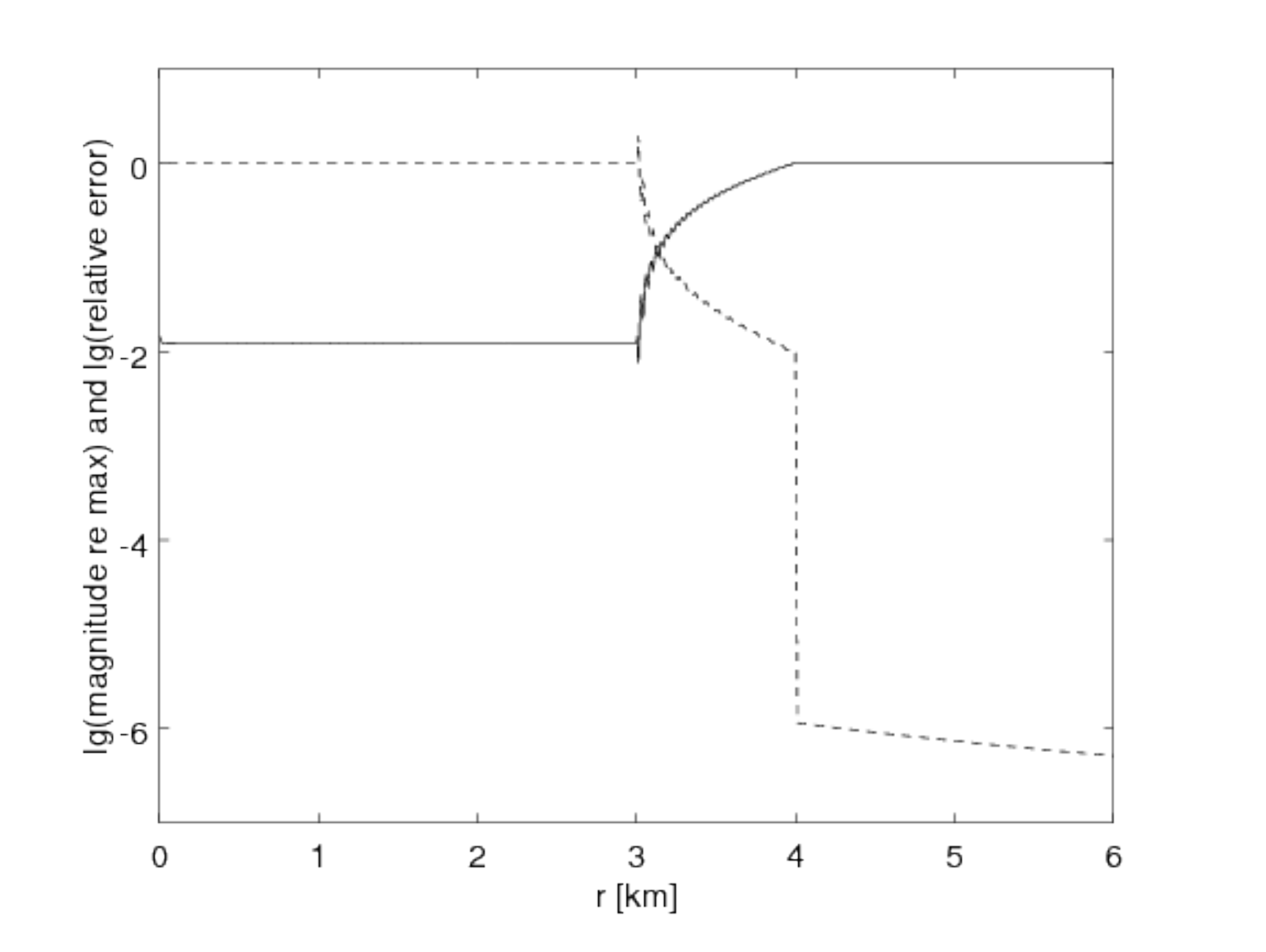}
\includegraphics [width=7.0cm] {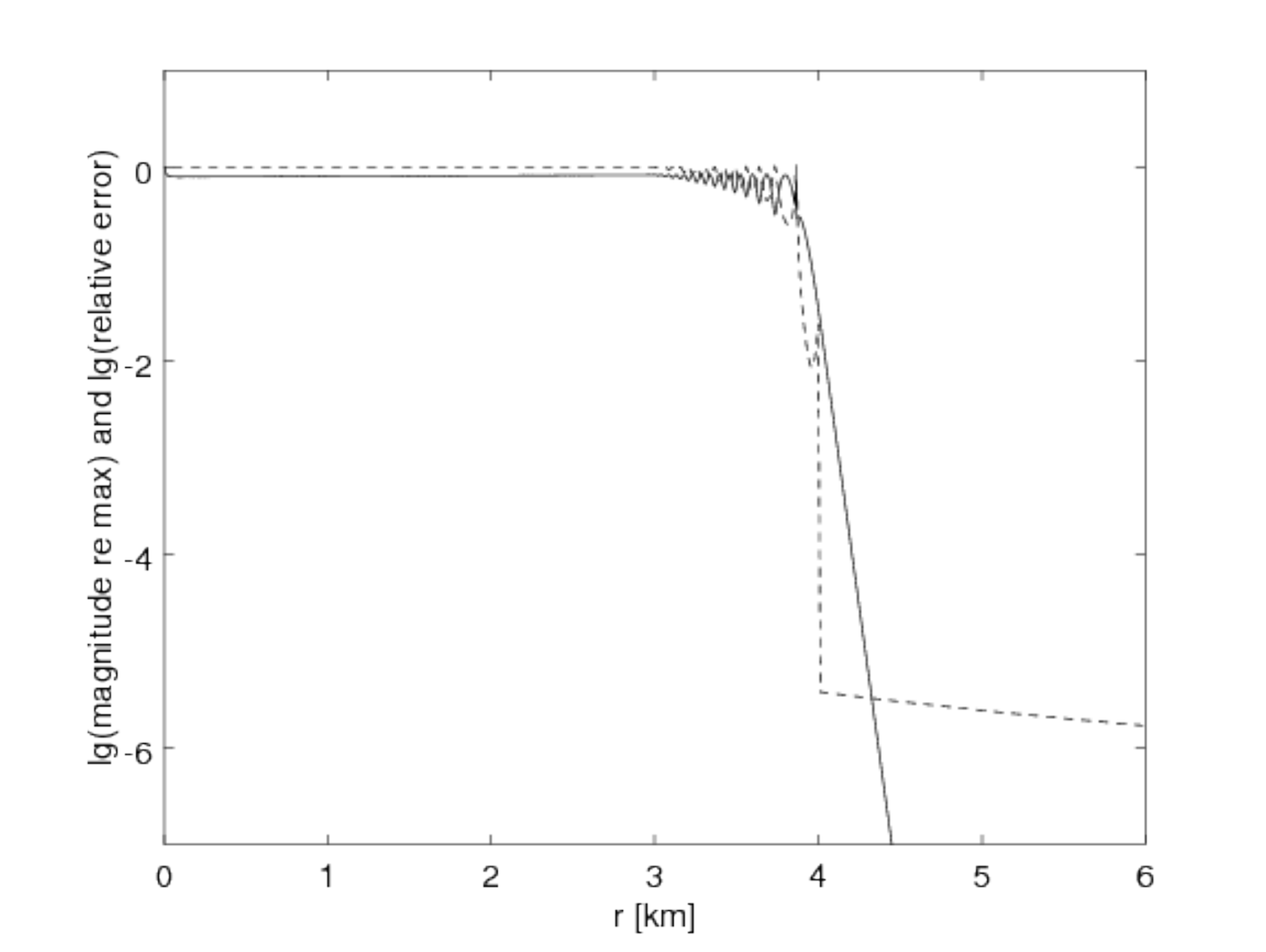}
\caption{ \textit{Solid curves}: Magnitude of the left-hand side
  of Eq.~(\ref{eq:pe2d_eta}), 
  with $ \gamma_{m} / H_{0}^{(1)}(k_{0}r)$ replacing $\eta_{m}$, 
  for mode 5 (left panel) and 7 (right panel) in the 2-D example. 
  The curve shows the logarithm of the magnitude divided by its
  maximum value. 
  \textit{Dashed curves}: Corresponding relative errors of
  the right-hand side of Eq.~(\ref{eq:pe2d_eta}). 
  The curve shows the logarithm of the relative error. }
\label{fig:rR6x2_5_7}
\end{center}
\end{figure}

With accurate computation of the $\gamma_{m}(r)$ from
Eq.~(\ref{eq:ansatz_2d}), and their derivatives, 
it becomes possible to assess the errors of 
Eq.~(\ref{eq:pe2d_eta}) and other PE approximations. 
Figure~\ref{fig:rR6x2_5_7} shows 
the magnitude of the left-hand side of Eq.~(\ref{eq:pe2d_eta}) for
modes 5 and 7, for the 2-D example
from Sec.~2.1. 
The relative errors of the right-hand side, compared to 
the left-hand side, are also included, as dashed curves. 
In each case, the reference wavenumber $k_{0}$ equals 
the pertinent modal wavenumber $k_{m,0}$,
and $ \gamma_{m}(r) / H_{0}^{(1)}(k_{0}r) $ replaces $\eta_{m}(r)$
at the evaluation of the two sides of Eq.~(\ref{eq:pe2d_eta}). 

%--------------------------------------------------------
%start Fig. rR6x2_5_7; 
%--------------------------------------------------------

The significantly different characters of modes 5 and 7 are 
apparent also from Fig.~\ref{fig:rR6x2_5_7}. 
At $r < 3$ km, where $k_{m}(r)$ is close to $k_{0}$, 
$ d \left( \gamma_{m}(r) / H_{0}^{(1)}(k_{0}r) \right) / d r $ 
would be very small if there were no back-scattered or reflected
field. 
At $r > 4$ km, where there is no in-coming field, a large 
$ d \left( \gamma_{m}(r) / H_{0}^{(1)}(k_{0}r) \right) / d r $ 
may arise from the mis-match between $k_{m}(r)$ and $k_{0}$. 
With weak back-scattering for mode 5, its 
$ d \left( \gamma_{m}(r) / H_{0}^{(1)}(k_{0}r) \right) / d r $ 
is almost one hundred times stronger at $r > 4$ km than
at $r < 3$ km. 
For mode 7, on the other hand, the strong back-scattering 
gives a large
$ d \left( \gamma_{m}(r) / H_{0}^{(1)}(k_{0}r) \right) / d r $ 
at $r < 3$ km, and the cutoff at about $r$ = 4 km gives
rise to an exponential decay beyond that range, also for 
$ d \left( \gamma_{m}(r) / H_{0}^{(1)}(k_{0}r) \right) / d r $. 

Turning to the relative PE approximation errors, they are of
course very small in the out-going field regime beyond
$r$ = 4 km, for both modes, and they decrease as the
asymptotic approximation of the Hankel function improves
with increasing range. 
At $r < $ 3 km, they are quite large, about 1 (100 \%). 
This is not serious for mode 5 with its weak
$ d \left( \gamma_{m}(r) / H_{0}^{(1)}(k_{0}r) \right) / d r $ 
there, implying very small absolute errors.
For mode 7 with its strong
$ d \left( \gamma_{m}(r) / H_{0}^{(1)}(k_{0}r) \right) / d r $ 
at $r < $ 3 km, however, this is clearly serious. 
A (one-way) PE model is not able to follow the 
standing-wave interference pattern for mode 7 in this case.

%------------------------------------------------------------------------------
\section{The 3-D case}
% se ``11/9 17'' blad 1,9, uace17 sec 2, ``5/8 16'', ``24/8 15''

From now on, consider particular media such that,
\textit{cf.}~\cite[Sec. 7.1.2]{brekh92}, 
\begin{equation} \label{eq:c0}
  \frac{1}{c^{2}(x,y,z)} = \frac{1}{c_{0}^{2}(z)} + S(x) \ , 
\end{equation}
where (for convenience) $S(x_{s})=0$ s$^{2}$/m$^{2}$. 
For simplicity, $ \rho(x,y,z) = \rho_{0}(z) $ depends only on the depth $z$,
and $S(x)$ is constant for large positive as well as large negative $x$. 
In particular, the medium parameters are independent of the $y$ coordinate. 

As in Sec.~2, there is a free boundary at depth $z=0$ and a free or rigid
boundary at depth $z_{h}$. The $Z_{m,0}(z)$, $m$ = 1,2,.., now form a
complete set of orthogonal local modes at $x=x_{s}$ with corresponding 
modal wavenumbers $k_{m,0}$ and depth integrals $I_{m,0}$ 
by Eq.~(\ref{eq:Idef}). 
The $Z_{m,0}(z)$ are local modes $ Z_{m}(x,y,z) = Z_{m,0}(z) $ at other
points $(x,y)$ in the horizontal plane too, with depth integrals 
$I_{m}(x,y) = I_{m,0}$ but with modal wavenumbers
$ k_{m}(x) = ( k_{m,0}^{2} + \omega^{2} S(x) )^{1/2} $. 
Expansion of the pressure $p$ in Eq.~(\ref{eq:helm}) in terms of
the $Z_{m,0}$, \textit{i.e.}, 
\begin{equation} \label{eq:ansatz}
  p(x,y,z) = \sum_{m=1}^{\infty} \gamma_{m}(x,y) \, Z_{m,0}(z)  \ , 
\end{equation}
provides, after multiplication with $\rho_{0}^{-1} \, Z_{n,0}$ and 
integration over depth $z$, the 2-D horizontal refraction
equations~\cite[Eq.~(3.163)]{ivan17a} (but with ``$(x_{s})$''
corrected to ``$(\bm{x}_{s})$'' in the right-hand side). 
The mode-separated Helmholtz-type horizontal refraction equations
for the modal expansion coefficients $\gamma_{m}(x,y)$, $m$ = 1,2,.., 
are exact for the particular media under consideration. 

Fourier transformation with respect to $y$ of the horizontal
refraction equations yields
\begin{equation} \label{eq:gamint}
  \gamma_{m}(x,y) = \int_{-\infty}^{\infty}
    \hat{\gamma}_{m}(x,\kappa) \exp(i \kappa y) \, d\kappa \ , 
\end{equation}
where the functions $\hat{\gamma}_{m}(x,\kappa)$ fulfil 
ordinary differential equations. 
The ordinary differential equations for 
$\hat{\gamma}_{m}(x,\kappa)$ can be solved conveniently by
2$\times$2 propagator-matrix techniques as
detailed in~\cite[Sec. 3.2.1.2]{ivan17a}, for example. 
Finally, adaptive integration with automatic
error control~\cite{ivan92} is convenient to compute the
wavenumber integral in Eq.~(\ref{eq:gamint}). 

It is apparent how to use Eq.~(\ref{eq:gamint}) for 
computation of field derivatives in connection
with Eq.~(\ref{eq:ansatz}). 
Computation of energy flux requires first-order field
derivatives. Specifically, the time-averaged energy-flux or
intensity vector $\bm{\Phi} (\bm{x})$ is,
by~\cite[Eqs.~(1-11.11b)]{pierce91} and
by~\cite[Eqs.~(1-8.12)]{pierce91},
\begin{equation} \label{eq:eflux}
  \bm{\Phi} (\bm{x}) = \frac{1}{2 \omega \rho(\bm{x})} 
      \mathrm{Im} \left( p^{*}(\bm{x}) \, \nabla p(\bm{x}) \right) \ , 
\end{equation}
where the asterisk denotes the complex conjugate.

%-------------------------------------------------------------------
\subsection{\textit{Horizontal rays}}
% se ``2/8 16''

Ray theory provides an approximate way to solve the 2-D
horizontal refraction equations, that
gives useful insight. The pertinent sound-speed profile
in the horizontal $xy$-plane is, for mode $m$ and the angular
frequency $\omega$, $c_{m}(x,\omega)$ as defined by 
\begin{equation} \label{eq:cm_def}
  \frac {1} {c_{m}^{2}(x,\omega)} =
     \left( \frac {k_{m,0}(\omega)} {\omega} \right)^{2} + S(x) \ , 
\end{equation}
emphasizing the frequency dependence in the notation. 
It is independent of $y$, and the ray tracing starts from $(x_{s},y_{s})$.
Let $i(x)$ denote the angle between a ray at $x$, before it possibly 
turns, and the $x$ axis. With $\sigma$ as the Snell parameter of the ray,
Snell's law gives 
\begin{equation} \label{eq:snell}
  \sin i(x) = \sigma \, c_{m}(x) = \frac {\sin i(x_{s})}
    { \left( 1 + (\frac{\omega}{k_{m,0}})^{2} \, S(x) \right)^{1/2}}
                                                                          \ , 
\end{equation}
since $S(x_{s})=0$ s$^{2}$/m$^{2}$. Hence, the ray turns when 
$ S(x) = - (k_{m,0}/\omega)^{2} \, \cos i(x_{s}) $. 
With the same ray directions $i(x_{s})$ at the source, a higher-order mode,
with a smaller $k_{m,0}$, does not penetrate as deep into a
high-velocity region as a lower-order mode does. 
When $S(x)$ is piece-wise linear, the rays are composed of parabolic segments
and traveltimes along the rays can be computed analytically. 

Consider horizontal rays of a certain type, concerning start direction
towards increasing or decreasing $x$, number of turns \textit{etc.}, 
from $(x_{s},y_{s})$ to a fixed end point $(x_{e},y_{e})$ for mode $m$. The
corresponding paths $\Gamma_{m}(\omega)$, 
Snell parameters $\sigma_{m}(\omega)$, and traveltimes $T_{m}(\omega)$ 
depend in general on the angular frequency $\omega$,
since $k_{m,0}(\omega)/\omega$ does so. Apparently,
\begin{equation} \label{eq:Tn_def}
  T_{m}(\omega) = \int_{\Gamma_{m}(\omega)} \frac{ds}{c_{m}(x,\omega)} 
\end{equation}
where $s$ is arc length. It follows from Fermat's principle that
\begin{equation} \label{eq:fermat}
  T_{m}(\omega + d\omega) =
    \int_{\Gamma_{m}(\omega)} \frac{ds}{c_{m}(x,\omega+d\omega)} + o(d\omega) \ .
\end{equation}
Hence, by differentiation of Eq.~(\ref{eq:cm_def}) with respect to $\omega$, 
\begin{eqnarray}
  \frac {dT_{m}(\omega)} {d\omega} & = & 
    \int_{\Gamma_{m}(\omega)} \frac {dc_{m}^{-1}(x,\omega)} {d\omega} \, ds 
                                                             \label{eq:dTn1} \\
  & = & \frac{k_{m,0}(\omega)}{\omega^{2}}
    \left( \frac{dk_{m,0}(\omega)}{d\omega} - \frac{k_{m,0}(\omega)}{\omega} \right)
    \, \int_{\Gamma_{m}(\omega)} c_{m}(x,\omega) \, ds \label{eq:dTn2} \\
  & = & \frac{k_{m,0}(\omega)}{\omega^{2}}
    \left( \frac{dk_{m,0}(\omega)}{d\omega} - \frac{k_{m,0}(\omega)}{\omega} \right)
    \, \frac{|y_{e}-y_{s}|}{\sigma_{m}(\omega)} \ .    \label{eq:dTn3}
\end{eqnarray}
Propagator-matrix techniques~\cite{ivan98}, for example, can be used to compute 
the group slownesses $dk_{m,0}(\omega)/d\omega$. 

According to the stationary-phase approximation, frequency components around
$\omega$ interfere constructively at the group traveltime $t_{m}(\omega)$
defined by 
\begin{eqnarray}
  t_{m}(\omega) & = & T_{m}(\omega) + \omega \, dT_{m}(\omega)/d\omega
                                                              \label{eq:tn1} \\
  & = & T_{m}(\omega) + \frac{k_{m,0}(\omega)}{\omega} 
    \left( \frac{dk_{m,0}(\omega)}{d\omega} - \frac{k_{m,0}(\omega)}{\omega} \right)
  \frac{|y_{e}-y_{s}|}{\sigma_{m}(\omega)} \ .     \label{eq:tn2}
\end{eqnarray}
The usual group-slowness result
$ t_{m}(\omega) = \left( (x_{e}-x_{s})^{2} + (y_{e}-y_{s})^{2} \right)^{1/2} 
  dk_{m,0}(\omega)/d\omega $ appears 
when $S(x)$ vanishes identically.

%-------------------------------------------------------------------
\subsection{\textit{Example mimicking a 3-D wedge}}
% se uace17 sec 4

An example mimicking a 3-D wedge appears by a modification of the 
example in Sec.~2.1. 
The same Pekeris wave-guide is present at the source position
$(x_{s},y_{s})$ = (0,0) m, and the 30-Hz source is still at depth
$z_{s}$ = 106 m.

Select a continuous $S(x)$,
according to Eq.~(\ref{eq:c0}), such that the ``water'' sound speed increases 
to 1600 m/s at $x$ = 2.8 km and 1700 m/s at $x \geq$ 3.6 km. 
Furthermore, $S(x)$ is constant in $x \leq -3.6$ km, and it is linear in
$-3.6$ km $< x <$ 2.8 km as well as in 2.8 km $< x <$ 3.6 km. 
It follows that the ``water'' and ``sediment'' sound speeds are 1395.3 
and 2355.0 m/s, respectively, at $x \leq -3.6$ km, and that the ``sediment''
sound speed increases to 4177.9 m/s at $x$ = 2.8 km and 8876.9 m/s at
$x \geq$ 3.6 km. 
Another choice of $S(x)$ could of course provide more realistic sound-speed 
values, but the present choice is preferable for the illustrations. 
Recall that the density $ \rho(x,y,z) = \rho_{0}(z) $ is assumed to be
independent of $x$ and $y$. 

\begin{figure}[t]
\begin{center}
\vspace*{-2.0cm}                             %%%
\includegraphics [width=7.0cm] {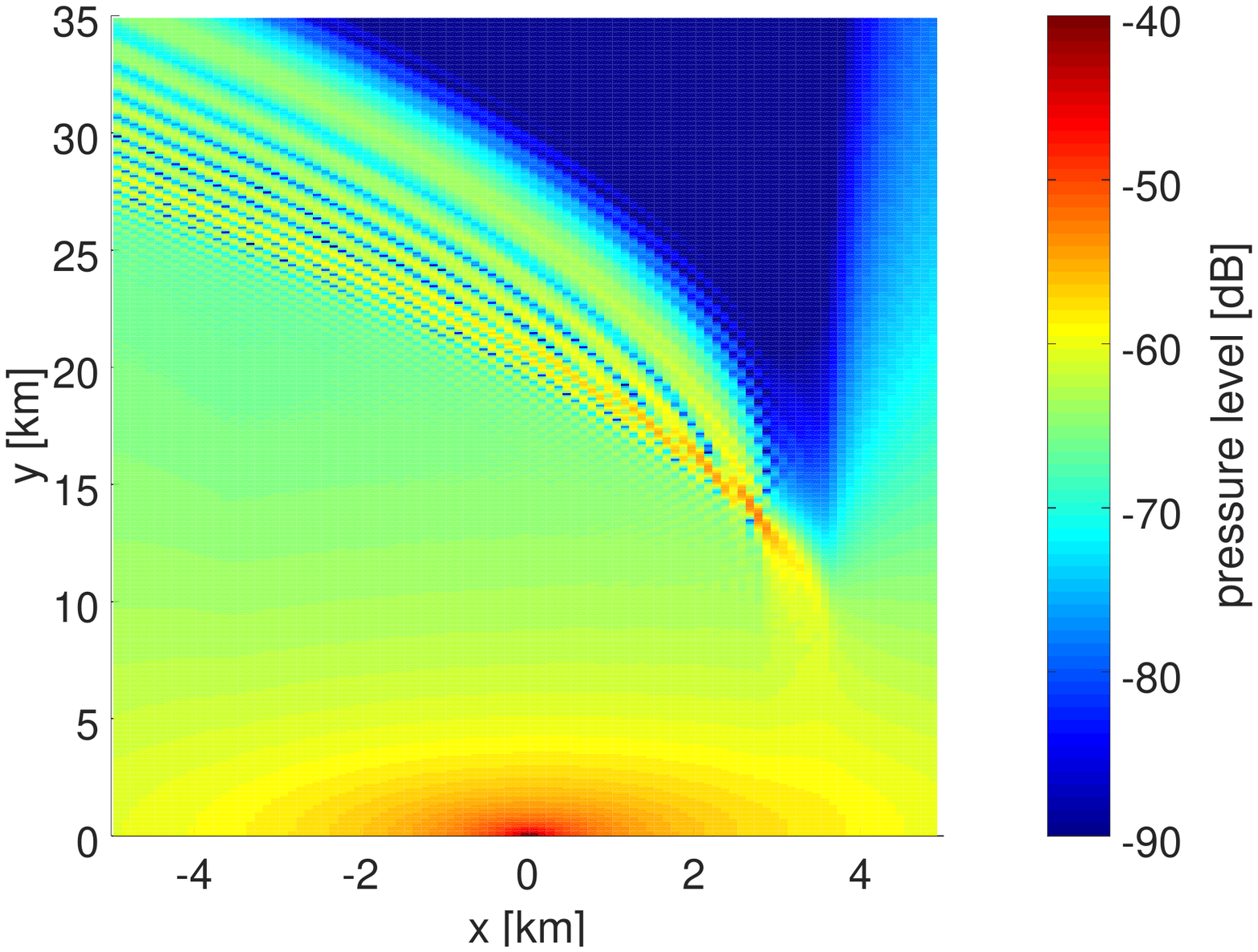}
\includegraphics [width=7.0cm] {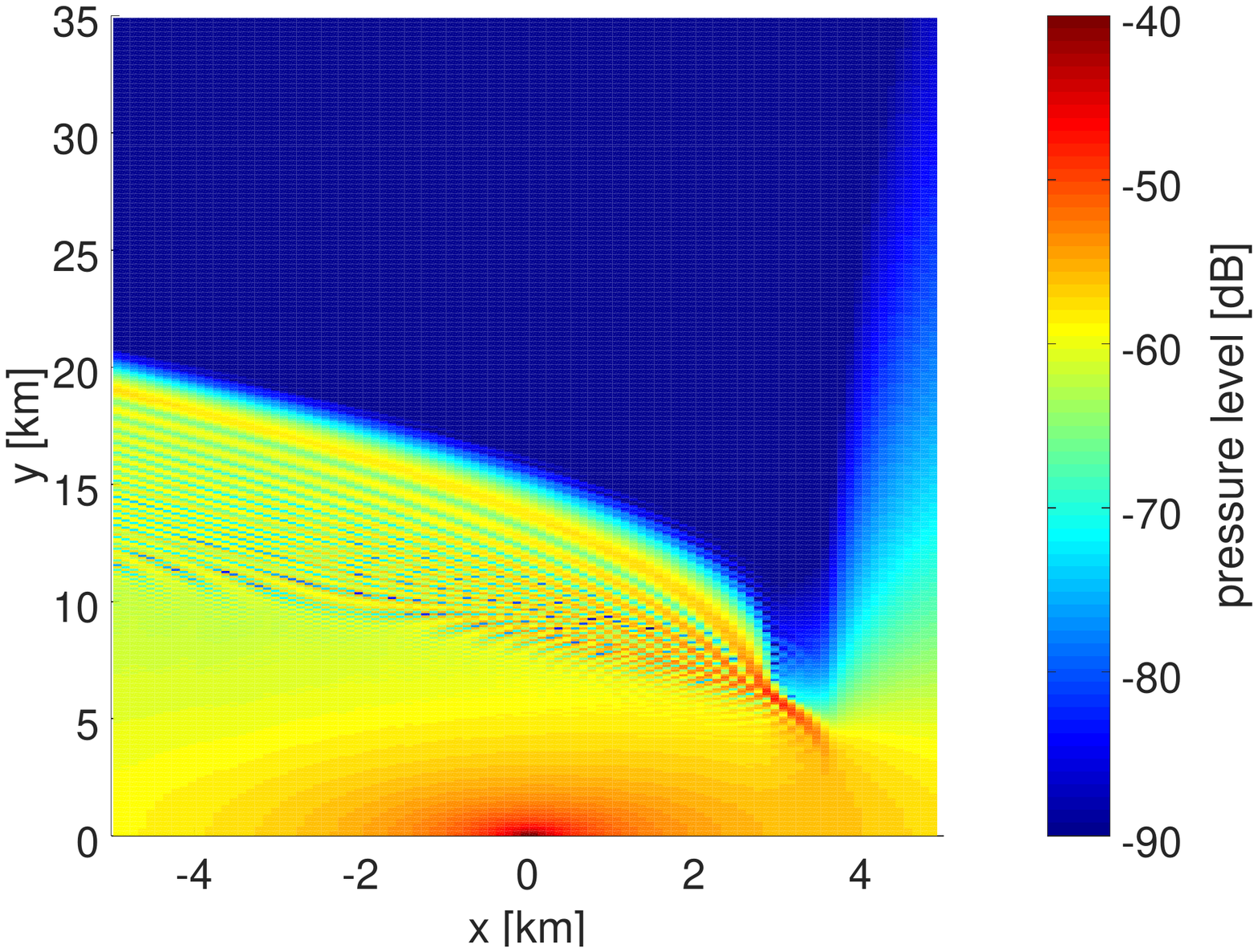}
\vspace*{-2.0cm}                             %%%
\caption{ Pressure level, in dB \textit{re}
  total spherical field at 1 m, of the field component given
  by mode 3 (left panel) and 7 (right panel), for the 3-D example, in the 
  horizontal $xy$-plane at depth $z$ = 106 m.
  The symmetric source is at $(x_{s},y_{s},z_{s})$ = (0,0,106) m
  and the frequency is 30 Hz. }
\label{fig:XY_3_7}
\end{center}
\end{figure}

\begin{figure}[t]
\begin{center}
\vspace*{-2.0cm}                             %%%
\includegraphics [width=7.0cm] {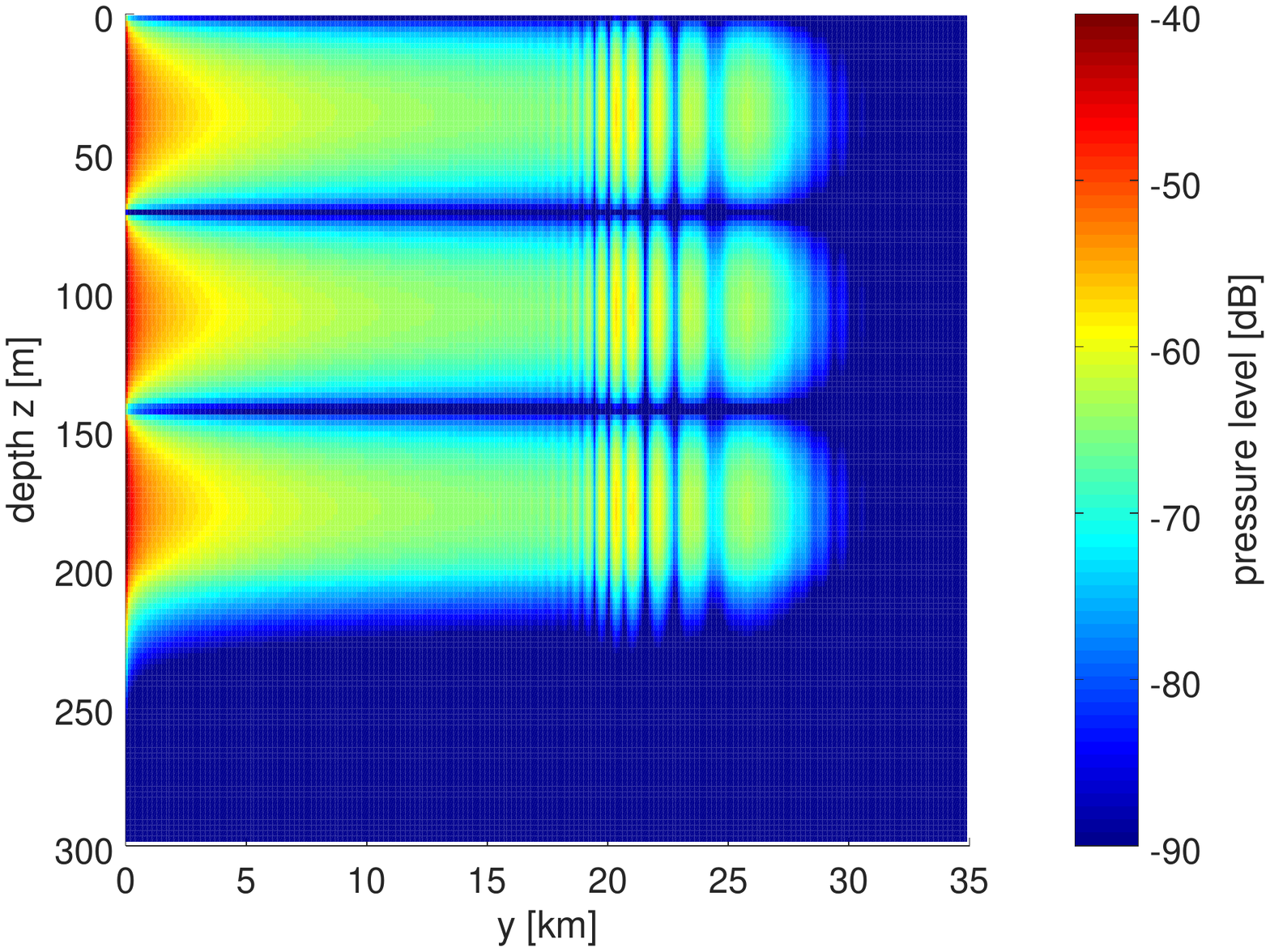}
\includegraphics [width=7.0cm] {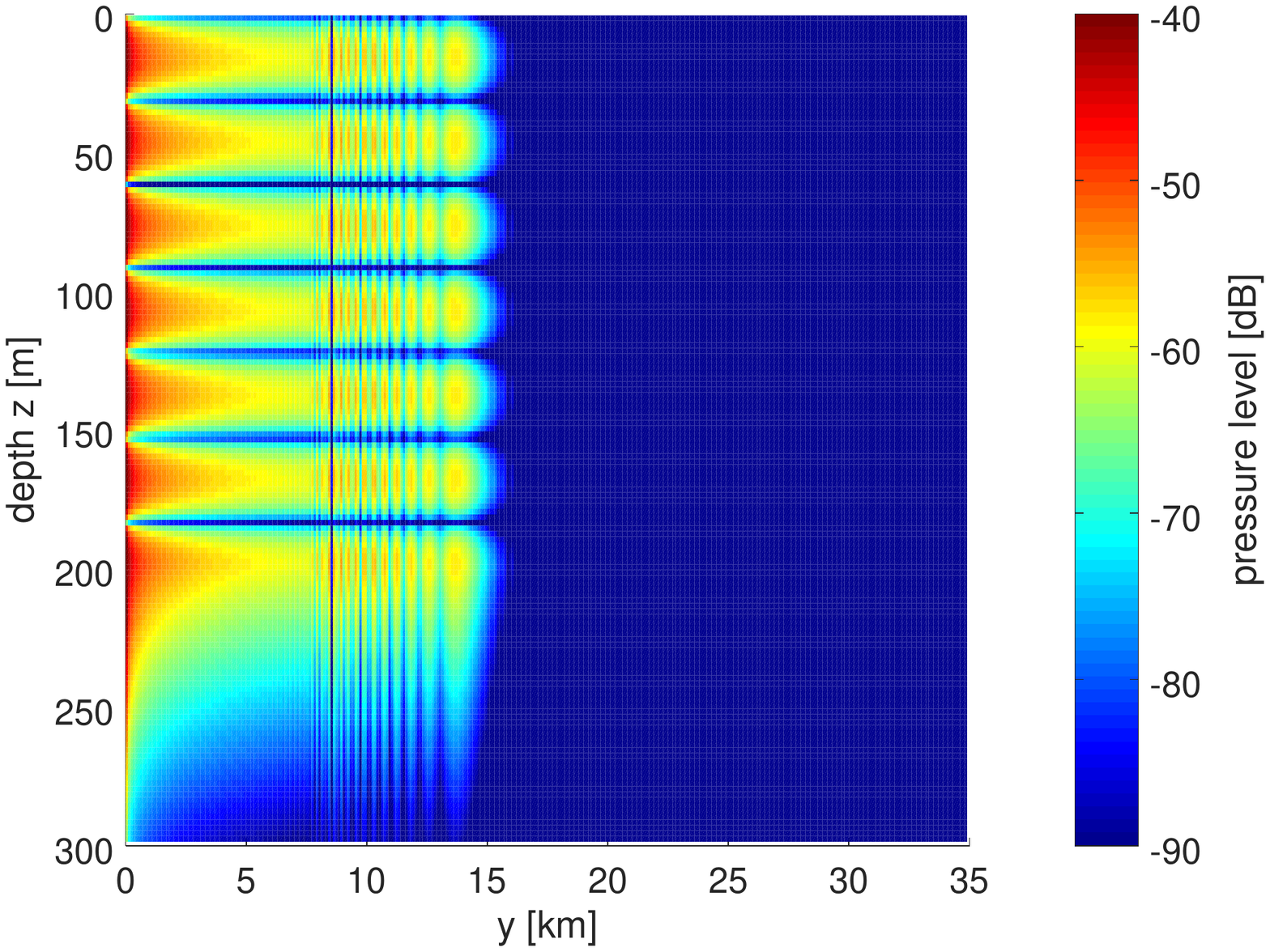}
\vspace*{-2.0cm}
\caption{ Pressure levels for modes 3 and 7 as in Fig.~\ref{fig:XY_3_7} 
  but in the $yz$-plane at $x$ = 0 m. }
\label{fig:YZ_3_7}
\end{center}
\end{figure}

Figure~\ref{fig:XY_3_7} shows pressure levels, in
dB \textit{re} total spherical field at 1 m, at depth $z$ = 106 m in the 
horizontal $xy$-plane for the mode 3 and mode 7 field components. 
The high-velocity region for positive $x$ causes clear effects of horizontal
refraction, particularly for the higher-order mode 7, 
\textit{cf.} Eq.~(\ref{eq:snell}). 
Depth dependencies of the field, as shown in Fig.~\ref{fig:YZ_3_7} for 
modes 3 and 7, are useful to verify the particular mode character.
At $x$ = 0 m, the mode 3 and 7 contributions drop off significantly beyond
about $y$ = 28 km and $y$ = 15 km, respectively. 
Indeed, Figs.~\ref{fig:XY_3_7} and~\ref{fig:YZ_3_7} exhibit similar features 
as~\cite[Figs. 9-10]{sturm05} with 3-D PE solutions for a
sloping-bottom wedge. 

%--------------------------------------------------------
%start Fig. XY_3_7; 
%--------------------------------------------------------

%--------------------------------------------------------
%start Fig. YZ_3_7; 
%--------------------------------------------------------

\begin{figure}[t]
\begin{center}
\includegraphics [width=7.0cm] {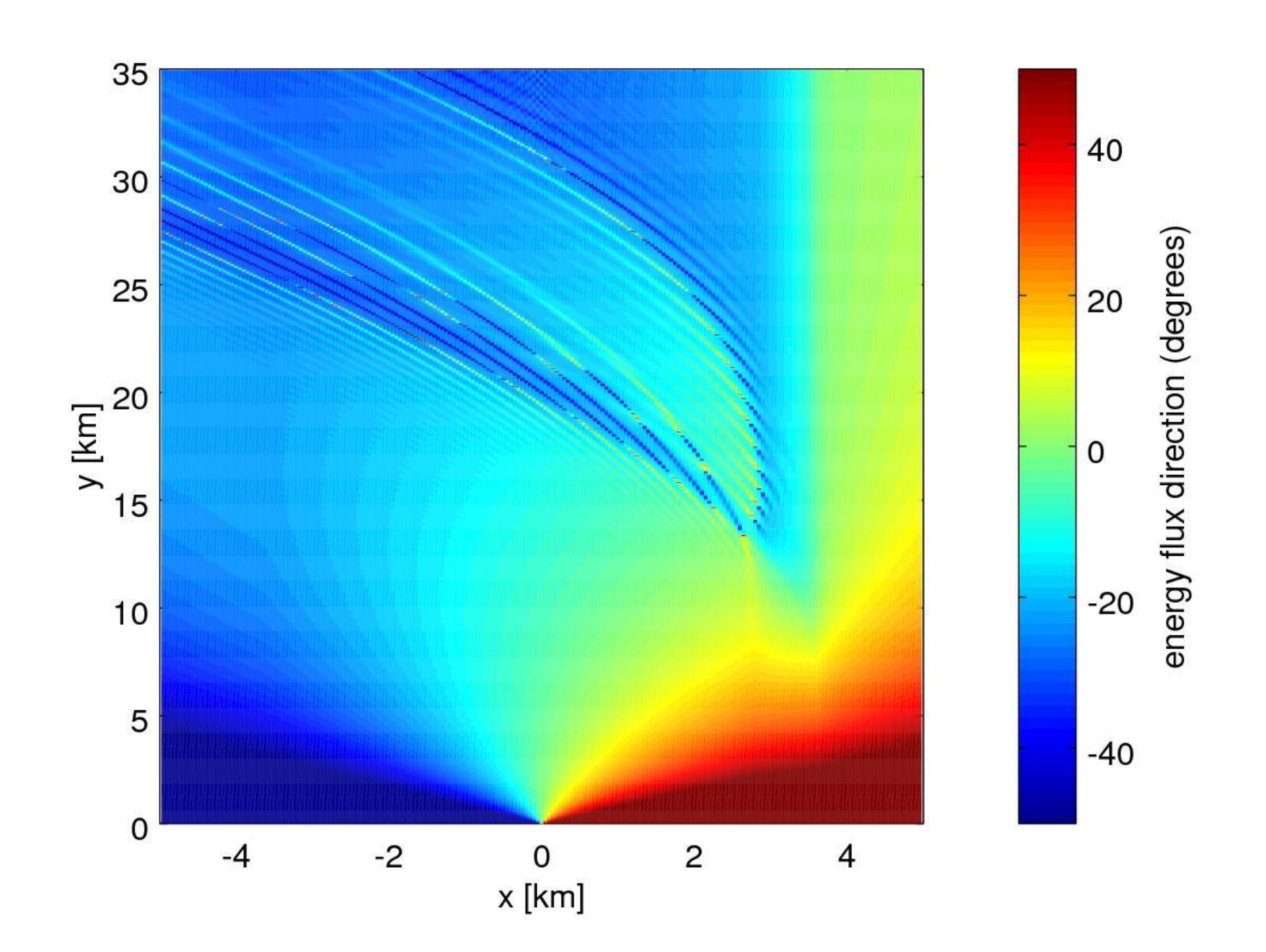}
\includegraphics [width=7.0cm] {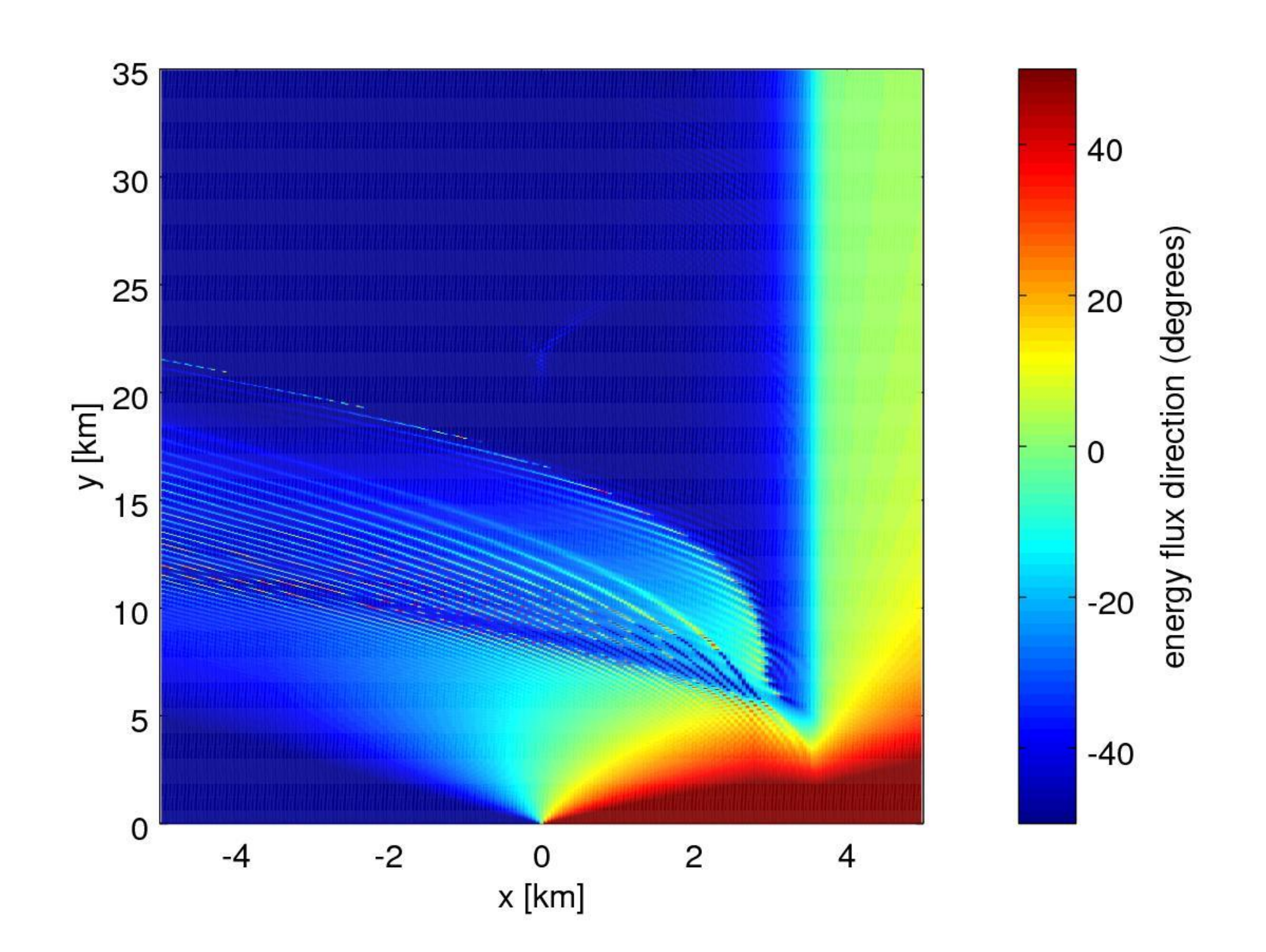}
%\vspace*{-2.0cm}
\caption{ Horizontal direction of the time-averaged energy flux of the
  field component given by mode 3 (left panel) and 7 (right panel),
  for the 3-D example, 
  in the horizontal $xy$-plane at depth $z$ = 106 m.
  Positive (negative) direction angles correspond to flux towards
  increasing (decreasing) $x$, and the angle 0$^{o}$ indicates
  flux in the $y$-axis direction. }
\label{fig:efluxXYarg50_3_7}
\end{center}
\end{figure}

The magnitudes of time-averaged energy flux for the mode 3 and mode 7 field 
components at depth 106 m, in dB \textit{re} plane-wave pressure
field of same strength as the total spherical field at 1 m, are rather
similar to the pressure-level results in Fig.~\ref{fig:XY_3_7}. 
Figure~\ref{fig:efluxXYarg50_3_7} shows the 
horizontal directions of the corresponding $\bm{\Phi} (\bm{x})$ vectors.
(Their vertical components are negligible at the source depth.) 
As expected, the energy radiates more or less radially close to
the source and at points $(x,y)$ with small $|y/x|$. 
The energy flux is in general directed to the left, towards 
decreasing $x$, in the shadow regions with weak 
fields, also in the right half-plane with $ x > x_{s} $ = 0 m. 
In addition, there are beams with energy flux to the left
starting at about $(x,y)$ = (3,13) km for the mode 3 field and
at about $(x,y)$ = (3,6) km for the mode 7 field. At $x = -5$ km, 
these beams reach $y \approx$ 28 and 12 km, respectively. 

%--------------------------------------------------------
%start Fig. efluxXYarg50_3_7; 
%--------------------------------------------------------

More ambitiously, it is possible to trace integral curves of a
vector field given by $\bm{\Phi} (\bm{x})$ according to
Eq.~(\ref{eq:eflux}). Foreman~\cite{foreman89} did this for some 
2-D cases, introducing the term ``exact rays'' for the integral
curves. In contrast to ordinary rays, such rays do not cross, since
there is a unique direction at each point. 

\begin{figure}[t]
\begin{center}
\vspace*{-2.0cm}
\includegraphics [width=7.0cm] {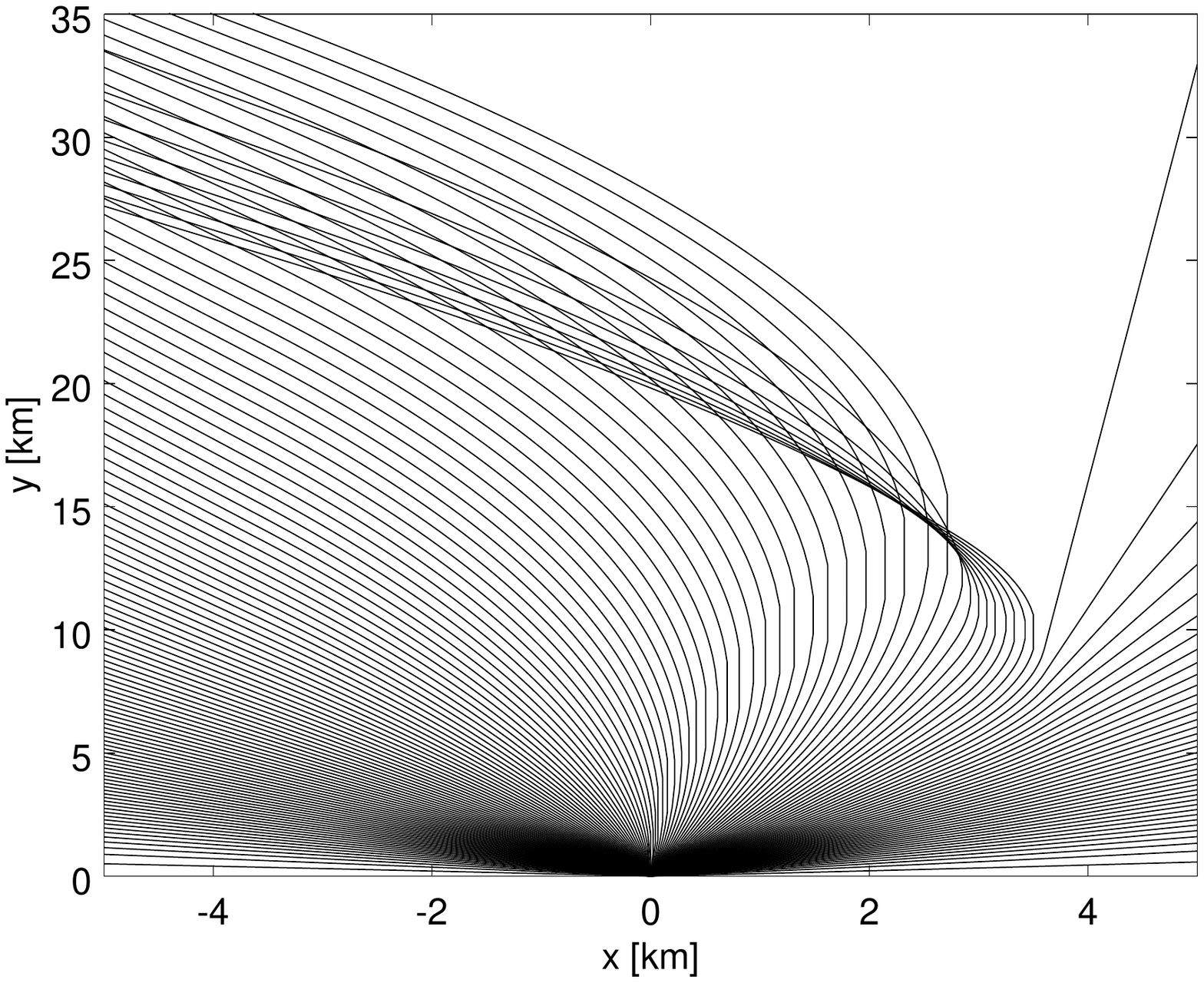}
\includegraphics [width=7.0cm] {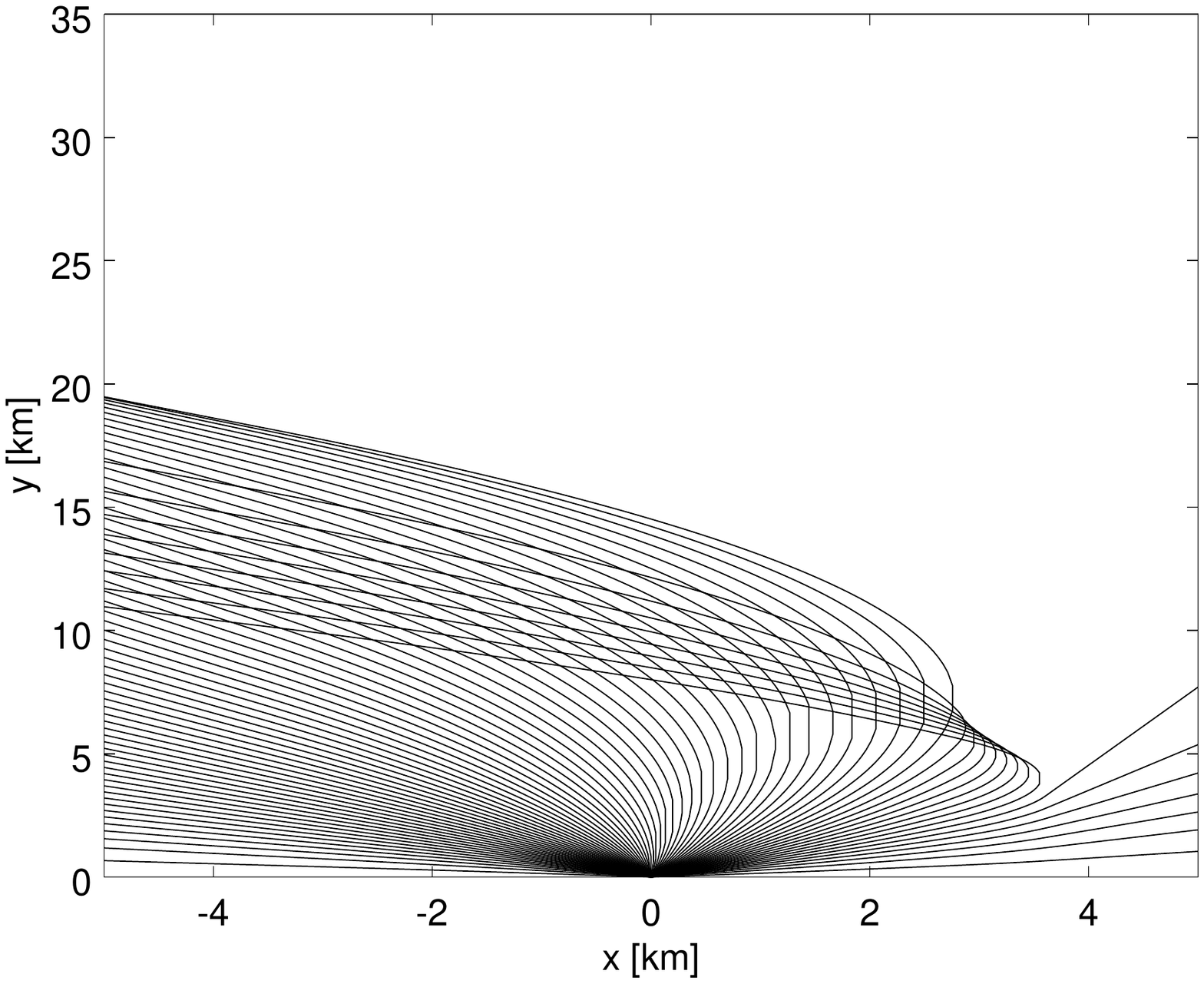}
\vspace*{-2.0cm}
\caption{ Ordinary horizontal rays for mode 3 (left panel)
  and 7 (right panel) at 30 Hz 
  according to the corresponding sound-speed profile $c_{m}(x,\omega)$ 
  as defined by Eq.~(\ref{eq:cm_def}). }
\label{fig:p99_3_7}
\end{center}
\end{figure}

The mentioned beams with energy flux directed to the left are
caused by horizontal refraction. 
Figure~\ref{fig:p99_3_7}, with ordinary
horizontal rays, provides further insight. The rays are traced
for 30-Hz sound-speed profiles
$c_{m}(x,\omega)$ with $m$ = 3 and 7, respectively, as 
defined by Eq.~(\ref{eq:cm_def}). 
In each case, there is a large-$y$ region with two ray arrivals 
to each point, separated from a shadow zone by a caustic ray 
envelope. At $x$ = 0 m, this region appears 
between about 20 and 28 km for mode 3, and 
between about 8 and 15 km for mode 7.
The interference between the two arrivals gives rise to 
Lloyd-mirror type patterns~\cite{jensen11} in 
Figs.~\ref{fig:XY_3_7} and~\ref{fig:YZ_3_7}. 
Moreover, the beams with energy flux directed to the left in 
Fig.~\ref{fig:efluxXYarg50_3_7} 
appear near the closest (smallest $y$) boundary of the
corresponding interference region, and they are connected
to ray turns at about $x$ = 3 km. 

%--------------------------------------------------------
%start Fig. p99_3_7; 
%--------------------------------------------------------

\begin{figure}[t]
\begin{center}
\vspace*{-2.0cm}
\includegraphics [width=7.0cm] {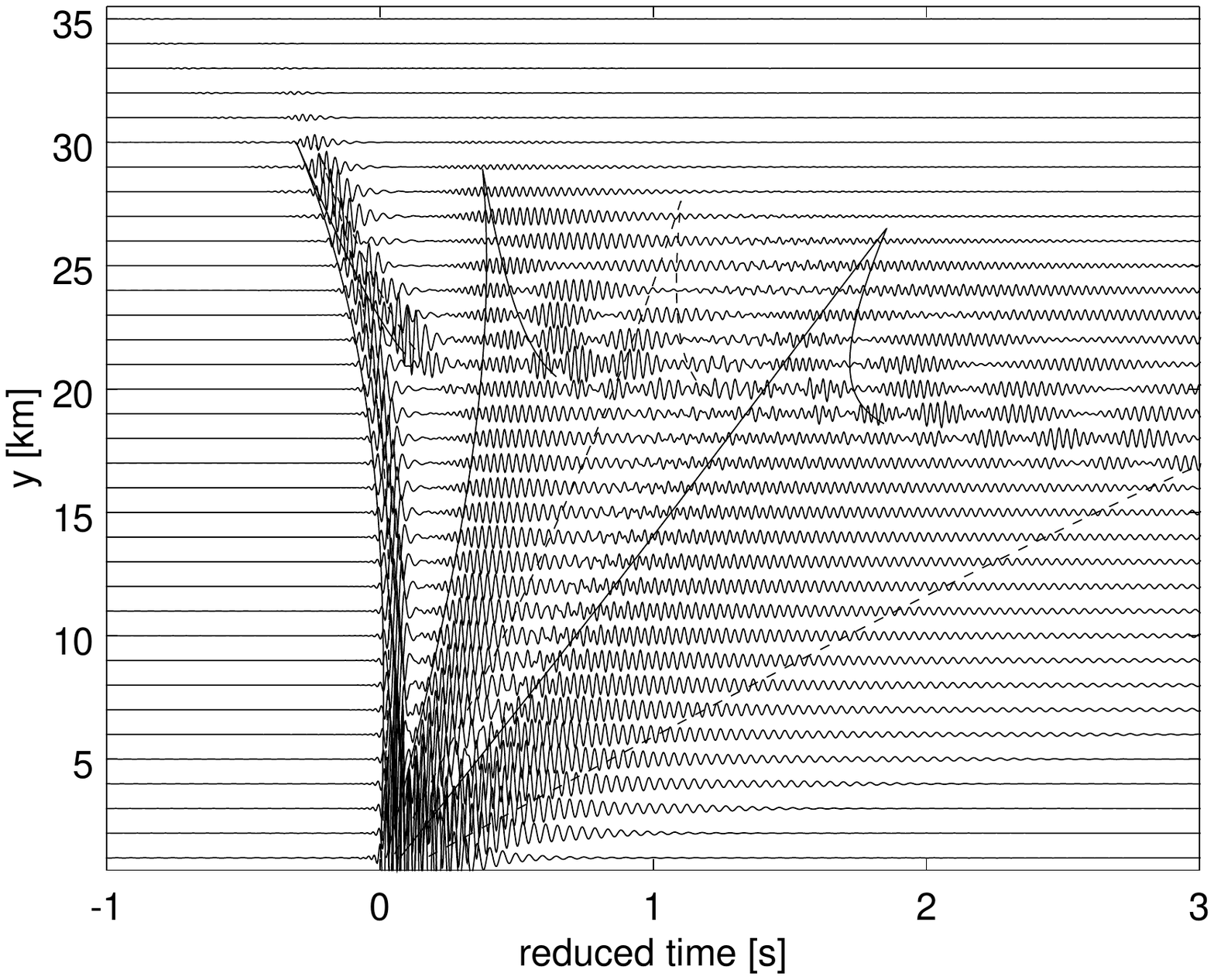}
\includegraphics [width=7.0cm] {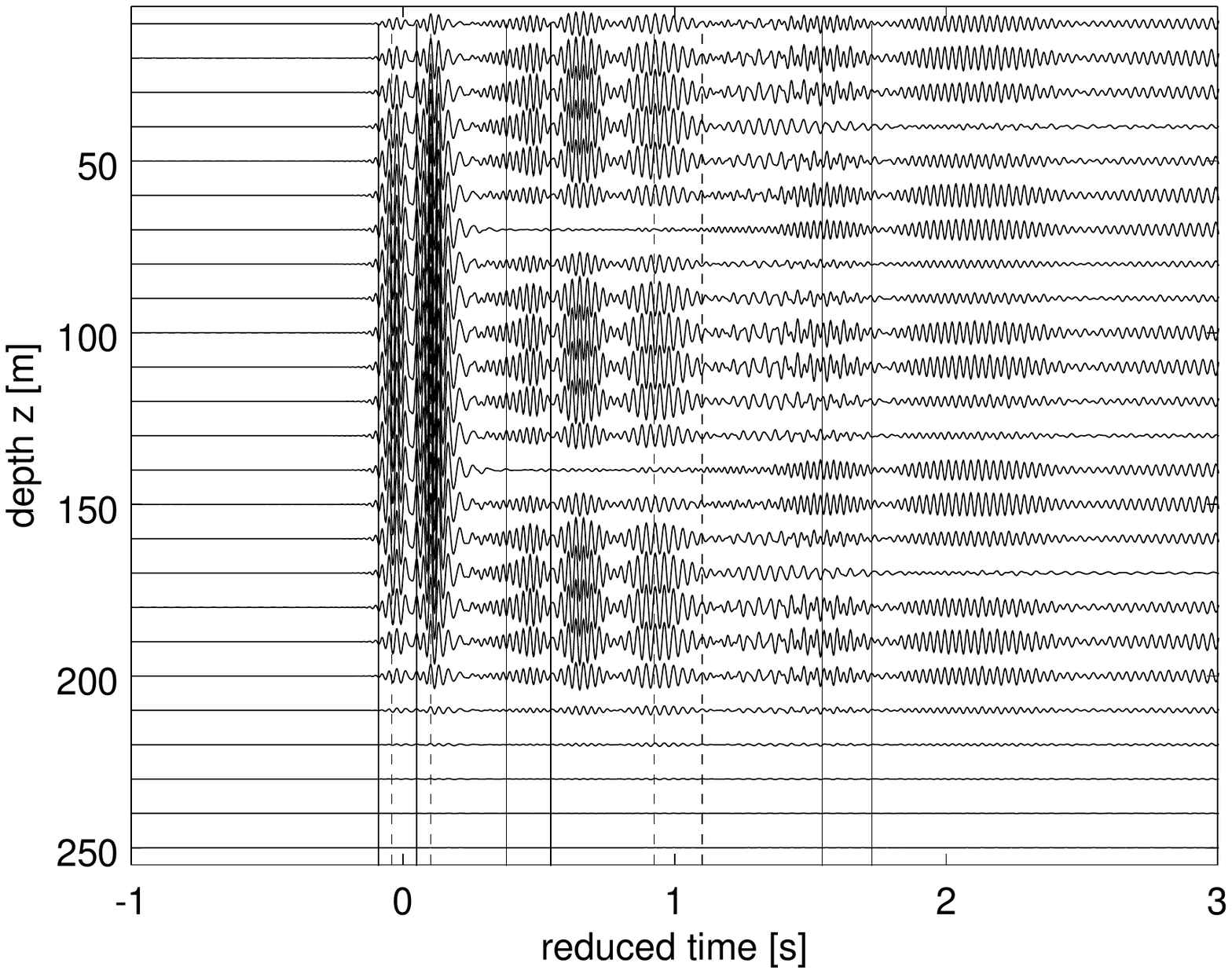}
\vspace*{-2.0cm}
\caption{ Time traces for the 3-D example with a short source pulse. 
  The horizontal axis indicates reduced traveltime $t - y/c$, with 
  $c$ = 1.5 km/s. 
  The additional solid and dashed curves show theoretical group 
  traveltimes at 40 and 28 Hz, respectively, for modes 1 (first),
  3 (next), and 5 (last). Mode 7, which would arrive later, is not
  included.
  The time traces are at $(x,z)$ = (0,106) m for $y$ = 1,2,..,35 km in
  the left panel, 
  and at $(x,y)$ = (0,22) km for $z$ = 10,20,..,250 m in the right panel. 
  (At $(x,y)$ = (0,22) km, the theoretical reduced group traveltimes for
  mode 5 at 28 Hz are larger than 3 s.) }
\label{fig:p22s_Zs}
\end{center}
\end{figure}

It is also instructive to apply Fourier synthesis to produce
the broad-band time traces in Fig.~\ref{fig:p22s_Zs},
for a short source pulse concentrated
to the frequency band (3-dB limits) 28-40 Hz. 
(Because of the presence of lower frequency components, the 
truncation boundary at depth 1450 m is here lowered to 3450 m.) 
The additional solid (for 40 Hz) and dashed (for 28 Hz) curves 
show theoretical group traveltimes $t_{m}(\omega)$, computed by 
Eq.~(\ref{eq:tn2}) for $m$ = 1, 3, 5. 
Modes 3 and 5 are apparently highly dispersive, and
around $y$ = 22 km (selected for the right panel of Fig.~\ref{fig:p22s_Zs}), 
there are two arrivals for each of modes 1, 3, and 5. 
For modes 3 and 5, the dispersion causes a
frequency beat pattern, since the instantaneous
frequencies of the two arrivals may differ by a few Hz
when they interfere. At some $y$ ranges, this gives
the false impression that there are more than two
arrivals for these modes. 

%--------------------------------------------------------
%start Fig. p22s_Zs; 
%--------------------------------------------------------

%-------------------------------------------------------------------
\subsection{\textit{3-D PE approximation errors}}
% se ``17/7 17'' blad 2-4, ``29/8 17'', ``18/7 16''

Assuming $(x_{s},y_{s})$ = (0,0) m, return to the 2-D PE 
given by Eq.~(\ref{eq:pe2d}). 
In the 3-D context, it represents an N~$\times$~2-D PE approximation,
ignoring horizontal refraction~\cite{jensen11}. 
With laterally invariant density, sound-speed variation according
to Eq.~(\ref{eq:c0}), local modes $Z_{m,0}(z)$ with modal wavenumbers 
$ k_{m}(x) = ( k_{m,0}^{2} + \omega^{2} S(x) )^{1/2} $, and
mode expansion according to 
$ \psi(r,\theta,z) = \sum_{m=1}^{\infty} \eta_{m}(r,\theta) \, Z_{m,0}(z) $, 
\textit{cf.} Eq.~(\ref{eq:ansatz}), 
the N~$\times$~2-D PE~(\ref{eq:pe2d}) takes the form 
\begin{equation}   \label{eq:pe3d_eta}
  \frac { \partial ( (k_{m}/k_{0})^{1/2} \, \eta_{m} ) } 
       { \partial r }  =  
    i k_{0} \, ( k_{m}/k_{0} - 1 ) 
       (k_{m}/k_{0})^{1/2} \, \eta_{m}
\end{equation}
for $m$ = 1,2,.., \textit{cf.} Eq.~(\ref{eq:pe2d_eta}). 
Using Fourier transformation according to Eq.~(\ref{eq:gamint}) and
wavenumber integration, it is now apparent how to evaluate the 
left- and right-hand sides of Eq.~(\ref{eq:pe3d_eta}), 
and variants thereof, with 
$ \gamma_{m} / H_{0}^{(1)}(k_{0}r)$ replacing $\eta_{m}$. 

\begin{figure}[t]
\begin{center}
\vspace*{-2.0cm}
\includegraphics [width=7.0cm] {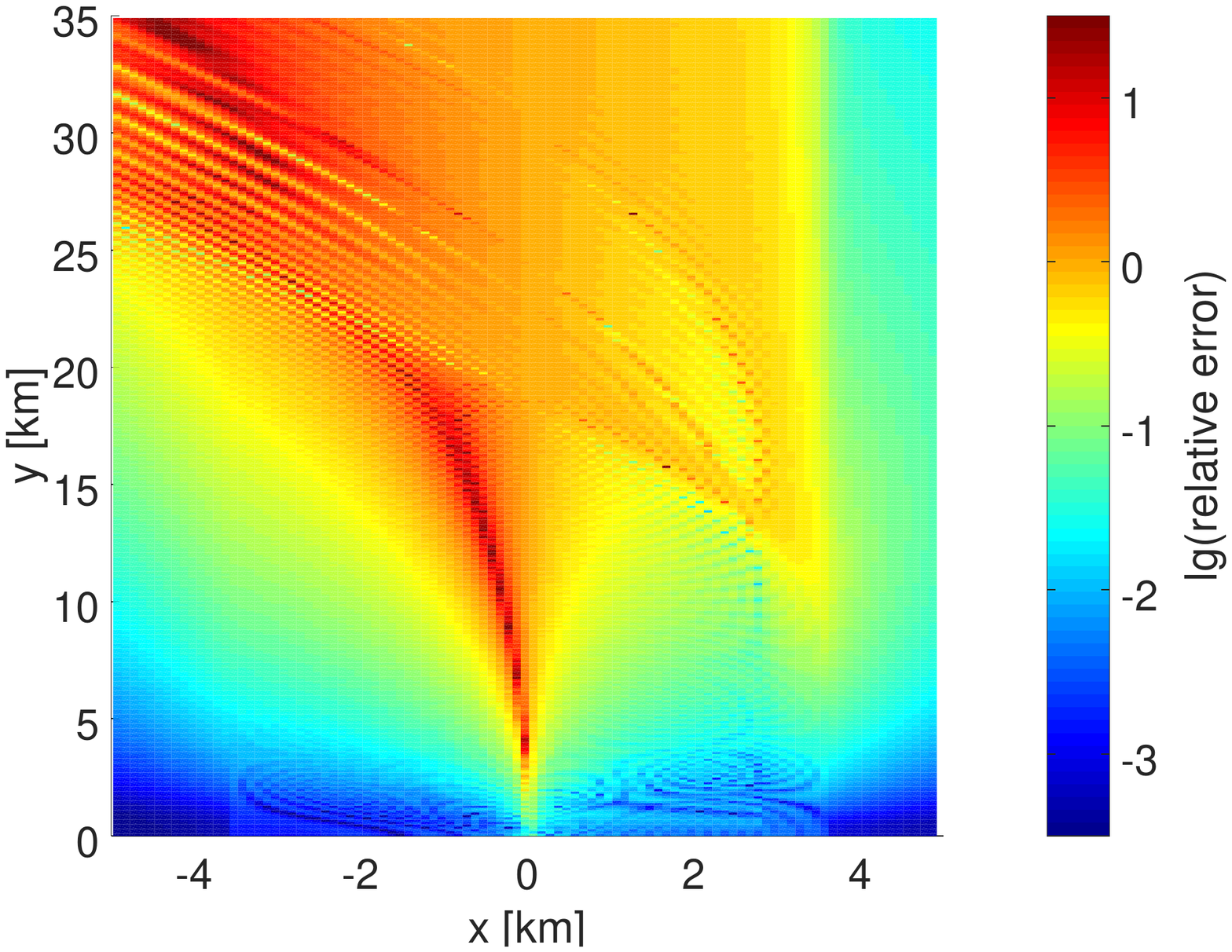}
\includegraphics [width=7.0cm] {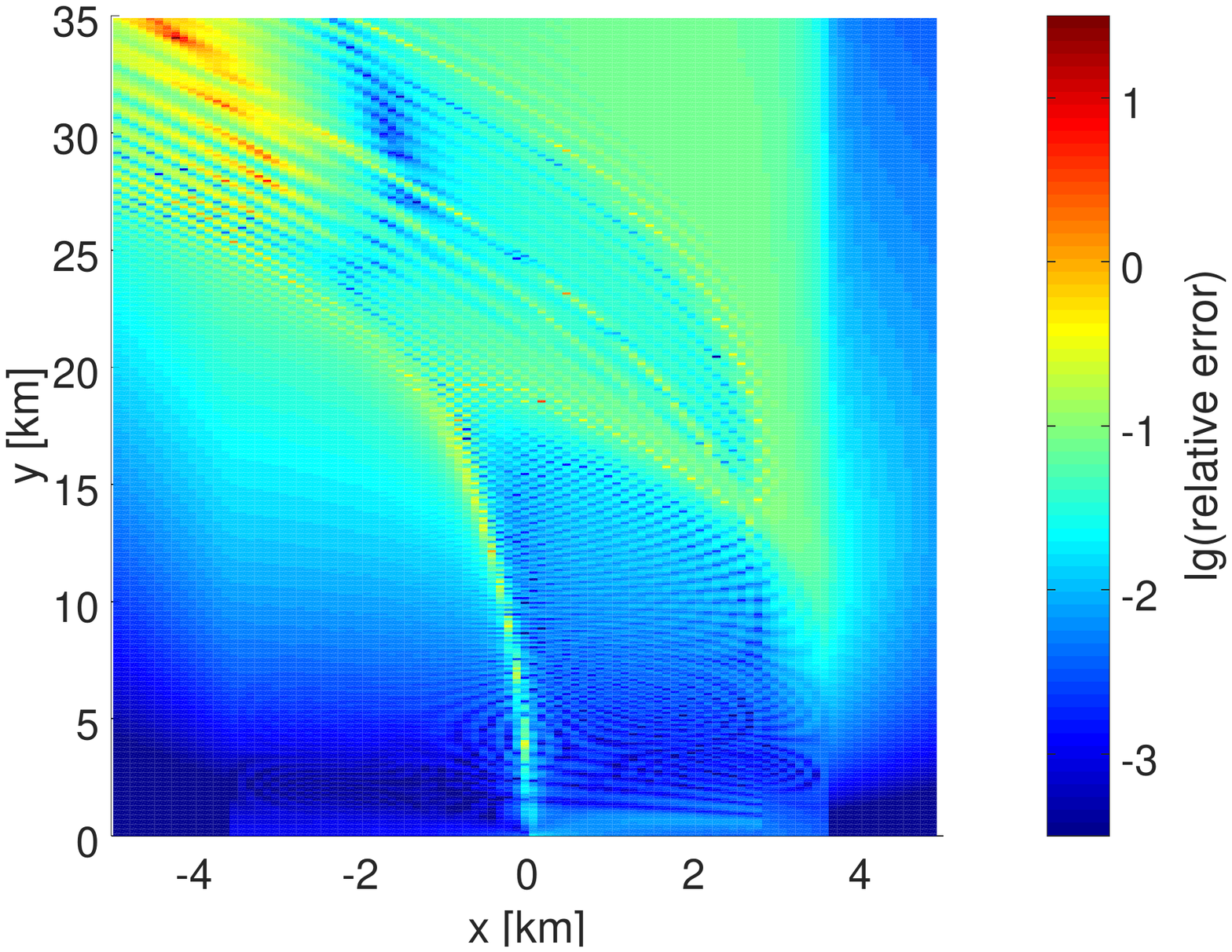}
\vspace*{-2.0cm}
\caption{ \textit{Left panel}: Relative approximation errors of the 
  N~$\times$~2-D PE~(\ref{eq:pe3d_eta}) 
  with $ \gamma_{m} / H_{0}^{(1)}(k_{0}r)$ replacing $\eta_{m}$, 
  for mode 3 in the 3-D example. 
  The map shows the logarithm of the relative error. 
  \textit{Right panel}: Relative approximation errors of a certain 
  3-D PE. }
\label{fig:rXY0a2bE_3}
\end{center}
\end{figure}

The left panel of Fig.~\ref{fig:rXY0a2bE_3} shows the relative errors of the 
right-hand side of Eq.~(\ref{eq:pe3d_eta}) compared to the left-hand side, for
the example from Sec.~3.2 and mode $m$ = 3. 
The reference wavenumber $k_{0}$ equals the
corresponding modal wavenumber $k_{m,0}$. 

%--------------------------------------------------------
%start Fig. rXY0a2bE_3; 
%--------------------------------------------------------

Effects of horizontal refraction can be incorporated by amending 
the N~$\times$~2-D PE~(\ref{eq:pe2d}) with a term involving a 
Pad\'{e} approximation of the operator
$ (I+Y)^{1/2} - I $, where 
$ Y = (k_{0} r)^{-2} \, \partial^{2} / \partial \theta^{2} $~\cite{sturm05}. 
The two sides of the resulting 3-D PE may be multiplied by the involved
Pad\'{e}-approximation denominators. 
Significantly reduced relative errors appear in this way, 
as shown by the right panel of Fig.~\ref{fig:rXY0a2bE_3}. 
For the part of the $xy$-plane that is shown, the area
fraction where the relative PE approximation errors are greater than 0.1
decreases from 68 \% for the left panel of Fig.~\ref{fig:rXY0a2bE_3} to
6 \% for the right panel. 

However, this 3-D PE is not always accurate enough 
and further improved 3-D PE approximations are
of interest. The Chisholm rational approximant for
two variables could be useful in this context~\cite{lee17}.

%------------------------------------------------------------------------------
\section{Concluding remarks}
%Haer i concl: FOER IHOP 2-D och 3-D!!! (olika aspekter!)

The horizontal refraction equations for 
modal expansion coefficients are exact for the class of media
with laterally invariant density and simple sound-speed variation
according to Eq.~(\ref{eq:c0}), allowing separate (adiabatic) handling of
the modes. 
The local modes do not vary among different horizontal 
positions, only the local modal wavenumbers do. 
After Fourier transformation with respect to one of the
horizontal coordinates, wavenumber integration according to
Eq.~(\ref{eq:gamint}) provides accurate numerical 
solutions for each modal field and its spatial derivatives. 
With mild restrictions, there are explicit expressions
for the integrand in terms of Airy and exponential functions. 
Computations of energy flux and horizontal rays help to
understand the structure of the field, and the group traveltime
result~(\ref{eq:tn2}), for frequency-dependent ray paths, 
is thereby useful at broad-band applications. 

For the azimuthally symmetric 2-D variant, with simple
range dependence according to Eqs.~(\ref{eq:rho0_2d})
and~(\ref{eq:c0_2d}), there are explicit expressions
for the modal expansion coefficients in terms of Hankel and 
Airy functions, without any wavenumber integration. 
A restriction is that on each subinterval $(r_{1},r_{2})$,
for a subinterval division of the range axis, 
the factor $R(r)$ is constant and
the term $S(r)$ is either constant or of the form 
\begin{equation} \label{eq:s_expr}
  S(r) = \frac { (r_{2}-r) S(r_{1}) + (r-r_{1}) S(r_{2}) }
               {(r_{2}-r_{1})} 
  + \frac{1}{4 \omega^{2}}
    \left( \frac{1}{r_{1}}-\frac{1}{r} \right)
    \left( \frac{1}{r}-\frac{1}{r_{2}} \right) 
    \left( 1 + \frac{r}{r_{1}} + \frac{r}{r_{2}} \right) . 
\end{equation}
Thus, application of the explicit Airy-function expressions 
implies different $S(r)$ for different angular 
frequencies $\omega$. At broad-band computations, 
handled with Fourier synthesis, interpolation with short
subintervals can mitigate the differences. 
With significant sound-speed increase with range, mode
cutoff may appear (Sec.~2.1), much as at 
upslope propagation in a homogeneous water column. 
However, the corresponding modes are reflected back
towards the source rather than lost by penetration
into the bottom. 

For the media according to
Eqs.~(\ref{eq:rho0_2d})-~(\ref{eq:c0_2d}) or 
Eq.~(\ref{eq:c0}) with a laterally invariant density, 
the adiabatic-mode computation methods of Secs.~2 and~3 
generate accurate reference solutions for
PE-model verification. 
The important issue of mode coupling effects remains,
however. 
Sections~2.2 and~3.3 include analysis examples for PE
approximation errors. 
In this context, the errors concern the
difference between the two sides of the 
PE~(\ref{eq:pe2d_eta}) or~(\ref{eq:pe3d_eta}) 
at insertion of Hankel-function scaled solutions 
of the Helmholtz-type equations~(\ref{eq:hrefract_2d})
or the horizontal refraction equations, respectively. 
Significant improvements by full 3-D, rather than
N~$\times$~2-D, computations show up in Sec.~3.3. 

It is possible to solve the 2-D PE~(\ref{eq:pe2d_eta}) 
as well as the N~$\times$~2-D PE~(\ref{eq:pe3d_eta}) 
analytically. Considering the 2-D PE~(\ref{eq:pe2d_eta}), for
example, with a modal start solution at $r = r_{0} \geq 0$, 
where the medium is range-invariant in the
interval $[0,r_{0}]$, its solution  is 
\begin{equation} \label{eq:eta_solv}
  \eta_{m}(r) = \frac {R^{1/2}(r)} {R^{1/2}(r_{0})} 
    \left( \frac {k_{m}(r_{0})} {k_{m}(r)} \right)^{1/2} 
  \exp \left( i \int_{r_{0}}^{r} (k_{m}(u)-k_{0}) \, d u \right) 
                          \eta_{m}(r_{0})     
\end{equation}
for $m$ = 1,2,... 
Hence, it is easy to compare $\eta_{m}(r)$ and 
$ \gamma_{m}(r) / H_{0}^{(1)}(k_{0}r) $ analytically, 
where $\gamma_{m}(r)$ solves Eq.~(\ref{eq:hrefract_2d}). 
This is also true for variants of Eq.~(\ref{eq:pe2d_eta}), 
with Pad\'{e} approximations of the depth operators
$ (I+X)^{1/2} $ and $ (I+X)^{1/4} $ in Eq.~(\ref{eq:pe2d}). 
A few examples appear in~\cite[Sec. 3.4.4]{ivan17a}.

%------------------------------------------------------------------------------

\end{document}